\documentclass{aa}  
\usepackage{amstext,amsmath}
\usepackage{graphicx}
\usepackage{textcomp}
\usepackage{txfonts}
\usepackage{time}
\usepackage{subfig}
\usepackage{multicol}
\usepackage{color}
\usepackage{url}
\usepackage{pstricks} 
\usepackage{natbib,twoopt}
\usepackage[breaklinks=true,colorlinks=true,urlcolor=blue,citecolor=blue,linkcolor=blue]{hyperref} 
\bibpunct{(}{)}{;}{a}{}{,} 
\makeatletter
\newcommandtwoopt{\citeads}[3][][]{\href{http://adsabs.harvard.edu/abs/#3}%
{\def\hyper@linkstart##1##2{}%
\let\hyper@linkend\@empty\citealp[#1][#2]{#3}}}
\newcommandtwoopt{\citepads}[3][][]{\href{http://adsabs.harvard.edu/abs/#3}%
{\def\hyper@linkstart##1##2{}%
\let\hyper@linkend\@empty\citep[#1][#2]{#3}}}
\newcommandtwoopt{\citetads}[3][][]{\href{http://adsabs.harvard.edu/abs/#3}%
{\def\hyper@linkstart##1##2{}%
\let\hyper@linkend\@empty\citet[#1][#2]{#3}}}
\newcommandtwoopt{\citeyearads}[3][][]%
{\href{http://adsabs.harvard.edu/abs/#3}
{\def\hyper@linkstart##1##2{}%
\let\hyper@linkend\@empty\citeyear[#1][#2]{#3}}}
\makeatother

\newcommand{\varcsec}{.\hspace{-0.9mm}'\!\hskip0.4pt'\hspace{-0.2mm}} 

\usepackage{fixltx2e}
\usepackage{url}
\usepackage{epic}
\usepackage{rotating}
\usepackage{breqn}
\usepackage{multirow} 
\usepackage{array} 
\usepackage{balance}
\usepackage{epic}
\usepackage{xcolor}
\usepackage{xifthen}

\makeatletter
\def\url@leostyle{%
  \@ifundefined{selectfont}{\def\UrlFont{\sf}}{\def\UrlFont{\tiny\ttfamily}}}
\makeatother
\begin{document}
\authorrunning{T. Libbrecht et al. }

\title{Observations of Ellerman bomb emission features in He \textsc{i} D\textsubscript{3} and He~\textsc{i}~10830~\AA\,}

\author{Tine Libbrecht\inst{1} \and Jayant Joshi\inst{1} \and Jaime de la Cruz Rodr\'iguez\inst{1} \and Jorrit Leenaarts\inst{1} \and Andr\'es Asensio Ramos\inst{2,3}} 
\institute{Institute for Solar Physics, Dept. of Astronomy, Stockholm University, Albanova University Center, SE-10691 Stockholm, Sweden \email{tine.libbrecht@astro.su.se}
\and Instituto de Astrof\'{\i}sica de Canarias, 38205, La Laguna, Tenerife, Spain
\and Departamento de Astrof\'{\i}sica, Universidad de La Laguna, E-38205 La Laguna, Tenerife, Spain}

\date{Received <date> /
Accepted <date>}

\frenchspacing

\abstract
{Ellerman bombs (EBs) are short-lived emission features, characterized by extended wing emission in hydrogen Balmer lines. Until now, no distinct signature of EBs has been found in the He \textsc{i} 10830~\AA\,line, and conclusive observations of EBs in He \textsc{i} D\textsubscript{3} have never been reported.}
{We aim to study the signature of EBs in neutral helium triplet lines.}
{The observations consist of 10 consecutive SST/TRIPPEL raster scans close to the limb, featuring the H$\beta$, He \textsc{i} D\textsubscript{3} and He \textsc{i} 10830~\AA\,spectral regions. We also obtained raster scans with IRIS and make use of the SDO/AIA 1700 \AA\,channel. We use \textsc{Hazel} to invert the neutral helium triplet lines.}
{Three EBs in our data show distinct emission signatures in neutral helium triplet lines, most prominently visible in the He \textsc{i} D\textsubscript{3} line. The helium lines have two components: a broad and blue-shifted emission component associated with the EB, and a narrower absorption component formed in the overlying chromosphere. One of the EBs in our data shows evidence of strong velocity gradients in its emission component. The emission component of the other two EBs could be fitted using a constant slab. 
Our analysis hints towards thermal Doppler motions having a large contribution to the broadening for helium and IRIS lines. We conclude that the EBs must have high temperatures to exhibit emission signals in neutral helium triplet lines. An order of magnitude estimate places our observed EBs in the range of $T\sim 2\cdot 10^4-10^5$ K.}{}


\maketitle

\section{Introduction}\label{sec:intro}
In 1917, Ferdinand Ellerman published the first observations of what would later be called Ellerman bombs (\citeads{1917ApJ....46..298E}). He described the spectral features as the sudden appearance of narrow, bright bands in both wings of the H$\alpha$ line. A hundred years later, Ellerman bombs (hereafter EBs) still have an observational definition similar to the original description of Ellerman. EBs are transient events with lifetimes of several minutes and spatial scales of the order of 1\arcsec. Their most characteristic property is their Balmer line signature: extended wing brightenings and an unaffected line core due to an overlying optically thick layer of fibrils. A high-resolution demonstration of line core obscuration by fibrils is given in Fig.~1 of \citetads{2011ApJ...736...71W}. 

EBs occur in active regions and are usually associated with regions of enhanced magnetic activity and/or flux emergence. Observations of mixed polarities, flux cancellation and bi-directional jets at EB locations strongly hint to magnetic reconnection as the physical driver for EBs (e.g. \citeads{2002ApJ...575..506G,2004ApJ...614.1099P,2008ApJ...684..736W,2011ApJ...736...71W,2013ApJ...779..125N,2016ApJ...823..110R}). Recently, Quiet-Sun Ellerman-like Brightenings (QSEBs) have been discovered outside active regions (\citeads{2016arXiv160603675R}).
  
Several studies have used forward modelling to obtain temperature estimates for EBs. A localized temperature and/or density increase is added to a standard atmosphere and synthetic line profiles of H$\alpha$ (and sometimes Ca \textsc{ii} H or Ca \textsc{ii} 8542 \AA) are compared with observations. Temperature enhancements are found of 1500 K by \citetads{1983SoPh...87..135K}, 600–1300 K by \citetads{2006ApJ...643.1325F}, 5000 K by \citetads{2013A&A...557A.102B} and 4000 K by \citetads{2014A&A...567A.110B}. They all located EBs close to or slightly higher than the height of the temperature minimum in the (quiet) solar atmosphere. A two-cloud fitting method was proposed by \citetads{2014ApJ...792...13H} who estimated temperature enhancements for EBs of 400 - 1000 K. The lower cloud accounts for the wing emission and the upper cloud causes core absorption in the H$\alpha$ spectral profile of the EB.

\citetads{2014Sci...346C.315P} reported observations of hot pockets of gas of a temperature of $T\sim 8\cdot 10^4$ K in the photosphere, based on the observation of double peaked and extremely broad Si \textsc{iv} 1400 \AA\, profiles with the Interface Region Imaging Spectrograph (IRIS, \citeads{2014SoPh..289.2733D}). They conclude their paper by asking whether these IRIS bombs (IBs) could correspond to EBs. \citetads{2015ApJ...808..116J} independently analysed the same dataset and located these IBs higher up in the chromosphere. \citetads{2015ApJ...812...11V}, \citetads{2015ApJ...810...38K} and \citetads{2016ApJ...824...96T} combined ground based H$\alpha$ observations with IRIS observations and confirmed that EBs in certain cases can have strong signatures in IRIS lines Mg \textsc{ii} h\&k, C \textsc{ii} 1330 \AA\, and the Si \textsc{iv} doublet at 1400 \AA. Especially observations of broad and sometimes double peaked Si \textsc{iv} profiles are surprising since the coronal equilibrium formation temperature of the doublet peaks at $T = 8\cdot 10^4$ K. This result is in strong contradiction with all previously estimated temperature enhancements of EBs. 


EBs have observational signatures at multiple wavelengths: they show bright wings in Ca \textsc{ii} H\&K lines (e.g.  \citeads{1983SoPh...87..135K,2010PASJ...62..879H,2007A&A...473..279P}), and Ca \textsc{ii} 8542 \AA\;(e.g. \citeads{2006ApJ...643.1325F,2006SoPh..235...75S,2007A&A...473..279P,2013ApJ...774...32V,2014ApJ...792...13H,2015RAA....15.1513L}). EBs are also bright in mid-UV continua as seen in observations at 1600 and 1700 \AA\ (e.g. \citeads{2000ApJ...544L.157Q,2013ApJ...774...32V}). As discussed above, EBs can have signatures in IRIS lines Mg \textsc{ii} h\&k, Si \textsc{iv} 1400 \AA\, and C \textsc{ii} 1330 \AA. However, EBs are not visible in optical continua or in Na \textsc{i} D and Mg \textsc{i} b lines (\citeads{2015ApJ...808..133R}), nor in EUV emission lines as observed by the Atmospheric Imaging Assembly (AIA, \citeads{2012SoPh..275...17L}) on board of the Solar Dynamics Observatory (SDO).

In this paper, we study the signature of EBs in He \textsc{i} D\textsubscript{3} and He \textsc{i} 10830 \AA. In Ellermans original paper, \citetads{1909ApJ....30...75M} was cited, describing observations of an EB-like phenomenon exhibiting bright bands in the wings of He \textsc{i} D\textsubscript{3} as well as in H$\alpha$. Ellerman himself did not observe a He \textsc{i} D\textsubscript{3} brightening. To our knowledge, no other observations of EBs in He \textsc{i} D\textsubscript{3} have been published. \citetads{2015A&A...582A.104R} observed EBs in He \textsc{i} 10830 \AA, but only found co-spatial absorption with the EBs. However, using high-spatial resolution SST/TRIPPEL spectra, we have for the first time detected wing emission in He \textsc{i} D\textsubscript{3} and He \textsc{i} 10830~\AA\,associated with EBs. Hence, this paper adds another EB signature to the foregoing list of spectral lines and wavelength regions. 

EB signatures at various wavelengths are important because they provide constraints on modelling of EB atmospheres. As discussed, so far only H$\alpha$ and Ca \textsc{ii} lines have been modelled to estimate temperature enhancements. To add modelling of neutral helium triplet lines and IRIS Mg \textsc{ii} h\&k and Si \textsc{iv} lines, would put much stronger constrains to the problem. \citetads{2016A&A...590A.124R} claims that all combined visibilities of EBs (i.e. without helium) hold sufficient information to constrain temperatures of EBs between 10 000 -- 20 000~K and the hydrogen density to $\sim 10^{15}\,{\rm cm^{-3}}$. The question is now whether these temperature and density estimates are consistent with He \textsc{i} D\textsubscript{3} and He \textsc{i} 10830~\AA\,emission.

The commonly accepted line formation mechanism for He \textsc{i} D\textsubscript{3} and He \textsc{i} 10830~\AA\;in the upper chromosphere is through photoionization-recombination (\citeads{1939ApJ....89..673G,1975ApJ...199L..63Z,2005ApJ...619..604M,2008ApJ...681..650A,2008ApJ...677..742C,2016arXiv160800838L}). The neutral helium triplet levels are populated by incoming EUV radiation from the corona and transition region to the upper chromosphere, resulting in absorption lines formed by scattering of continuum photons. For EBs however, it seems counter intuitive that any influence from the corona or transition region would play a key role in the helium line formation of these deeply-located phenomena.

We proceed by describing the observations and data reduction in Sec.~\ref{sec:obs}. We discuss light curves and spectra of the EBs in Sec.~\ref{subs:det} and Sec.~\ref{subs:spec}. Inversion results using \textsc{Hazel} are shown in Sec.~\ref{subs:inv}. We compare EBs with emission signals to other EBs in our data in Sec.~\ref{subs:cond}. The results are put in a broader context and temperature estimates are given in Sec.~\ref{sec:disc}. Section~\ref{sec:concl} contains a summary and conclusions.

\section{Observations and data reduction}\label{sec:obs}

\subsection{SST Observations and Instrumentational Setup}
The observations were made with the Swedish 1-m Solar Telescope (SST, \citeads{2003SPIE.4853..341S}) on La Palma, employing the TRI-Port Polarimetric Echelle-Littrow spectrograph (TRIPPEL, \citeads{2011A&A...535A..14K}). 

\begin{table*}
\footnotesize
\caption{Overview of instrumentational parameters for the SST/TRIPPEL observations \label{instr}}
\begin{tabular}{c c c c c c c c c c}
\hline
\hline
& & & & & & & & & \\[-2mm]
Type & Wavelength & Camera & CCD size & Spectral & Binning & Spatial & Binning & Exposure & FOV \\[2mm]
 & \AA & & pix & \AA\,pix$^{-1}$ & \AA\,pix$^{-1}$ & \arcsec\,pix$^{-1}$ & \arcsec\,pix$^{-1}$ & ms & \arcsec\,\\[2mm]\hline
& & & & & & & & & \\[-2.5mm]
Spectrum & 5873.6 -- 5886.6 & MegaPlus II es4020& 2048$\times$2048 & 0.0065 & 0.0195 & 0.034 & 0.1 & 3$\times$100 & 30$\times$57 \\[2mm]
Spectrum &  4856.9 -- 4866.0 & MegaPlus II es1603 &1534$\times$1024 & 0.009 & 0.0137 & 0.043 & 0.1 & 3$\times$100 & 30$\times$41 \\[2mm]
Spectrum & 10818.4 -- 10834.9 & OWL SWIR 640\tablefootmark{2} & 640$\times$512 &  0.0264 & / & 0.067 & 0.1 & 3$\times$100 & 30$\times$34\\[2mm]
SJI & 5579 (44)\tablefootmark{1} & MegaPlus II es4020& 2048$\times$2048 & / & / & 0.034 & / & 3$\times$100 & 60$\times$60\\[2mm]
SJI & 10640 (30)\tablefootmark{1} & MegaPlus II es4020& 2048$\times$2048 & / & / & 0.034 & / &  3$\times$100 & 60$\times$60\\[2mm]
\hline
\end{tabular}
\tablefoottext{1} {The FWHM of the pre-filters are given between parenthesis.}\\
\tablefoottext{2} {Temporarily placed at the SST.}
\end{table*}

\begin{figure*}[!h]
\includegraphics[scale=1, width=\textwidth]{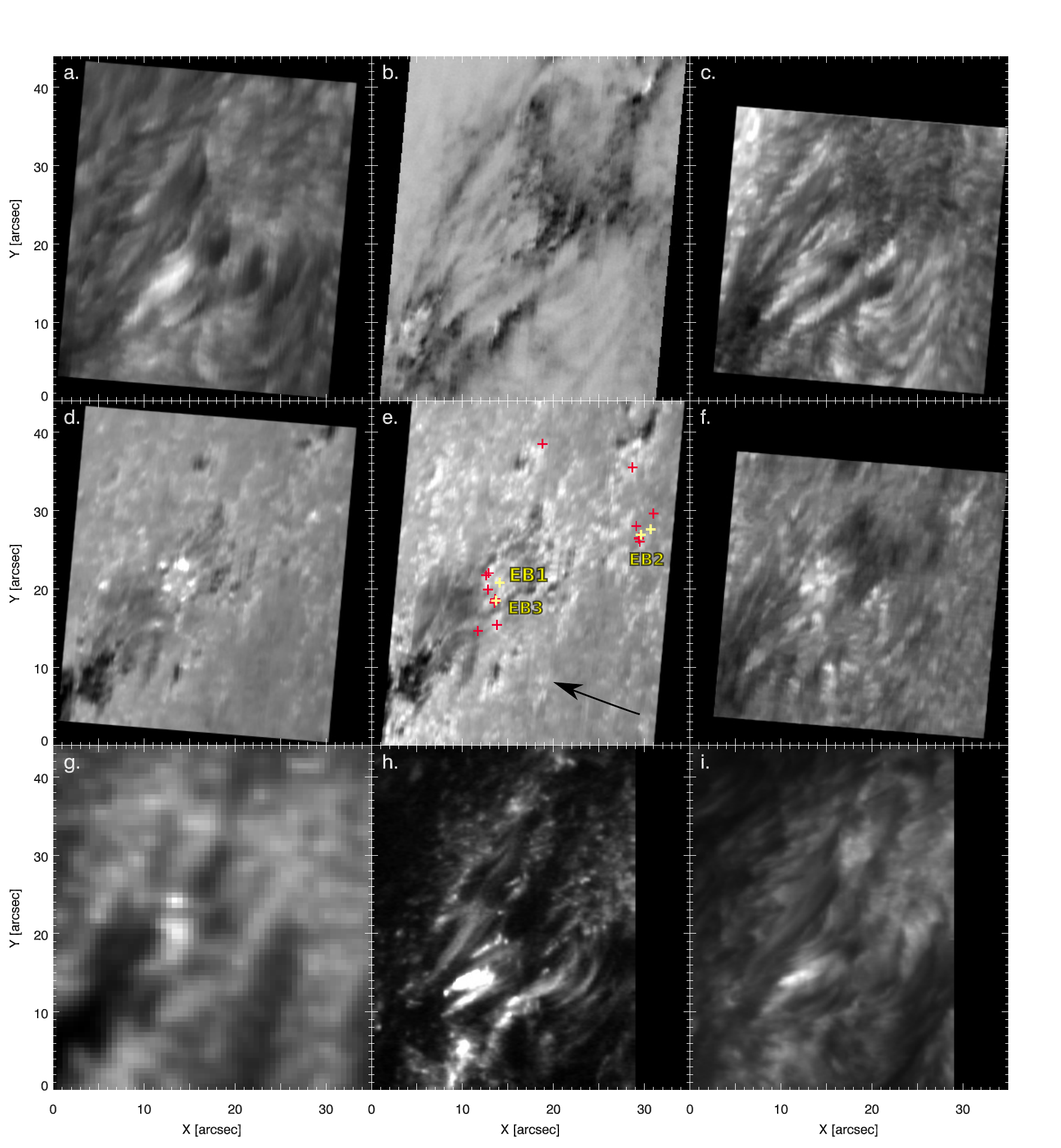}
\caption{\textbf{a}.-\textbf{f}. Raster scan images observed with SST/TRIPPEL: \textbf{a}. Line core raster scan in H$\beta$. \textbf{b}. Continuum-corrected line core raster scan in He \textsc{i} D\textsubscript{3}. \textbf{c}. Line core raster scan in He \textsc{i} 10830 \AA. \textbf{d}. Line wing raster scan in H$\beta-0.9$\AA. \textbf{e}. Line wing raster scan in He \textsc{i} D\textsubscript{3}+2.7\AA. Red plusses represent EB locations. Yellow plusses represent EB locations showing emission in He \textsc{i} D\textsubscript{3}, corresponding to EB1, EB2 and EB3. EB2 is indicated by two yellow plusses because it is split into two bombs EB2a and EB2b, corresponding to the same event at a different time step. The black arrow indicates the disk center direction. \textbf{f}. Line core raster scan in Si \textsc{i} 10827 \AA. \textbf{g}. SDO/AIA 1700 \AA\, image. \textbf{h}. Line core raster scan in Si \textsc{iv} 1400 \AA\,, observed with IRIS. \textbf{i}. Line core raster scan in Mg \textsc{ii} 2976 \AA, observed with IRIS. \label{panels}}
\end{figure*}

On 2015-08-01, we observed active region AR12390 at coordinates $x=870\arcsec,y=-268\arcsec$ and a viewing angle of $\mu=0.27$, close to the west limb. The active region appeared one week before our observations on 2015-07-24 and exhibited a C1.5 class flare on 2015-08-01 starting at 19:59 UT, 10h after our observations ended. The active region shows a complex topology, containing a sunspot, a pore with forming penumbra, several other pores and surrounding plage. In Fig.\,\ref{panels}, we show the raster scans for different spectral lines, together with context data from IRIS and SDO/AIA. 
The target is highly active as we observed several EBs and a transient loop brightening in the data. 

Between 07:51 UT and 10:10 UT, we obtained a total of 10 raster scans. Using three ports of TRIPPEL, we covered three spectral regions simultaneously, featuring H$\beta$ at 4861 \AA, He \textsc{i} D\textsubscript{3} at 5876 \AA, and He \textsc{i} 10830~\AA. No polarimetry was attempted. H$\beta$ and He \textsc{i} D\textsubscript{3} were recorded co-temporarily. He \textsc{i} 10830 \AA\;was recorded with a small time lag of less than 500 ms. This time difference is caused by the need of a focus shift between these spectral regions. In practice, the time difference between co-spatial points for H$\beta$, He \textsc{i} D\textsubscript{3} and He \textsc{i} 10830 \AA\;is larger due to differential atmospheric refraction. In the early morning at large airmass, differential refraction between the infrared and the optical can be of the order of several arcsecond, causing time differences in our observations up to 1 minute.

The correlation tracker of the SST was used to move the beam over the slit, taking 300 steps of $0\varcsec 1$, to obtain 30\arcsec\, wide raster scans. The projected slit is $0\varcsec 11$ wide. The total cadence for one raster scan of 300 steps is 11min 21s. This cadence includes 3 acquisitions with 100 ms exposure time for all spectral regions in each scanning position of the raster. Afterwards, the frame with the highest contrast has been selected. Table \ref{instr} summarizes instrumentational parameters of the SST observations.

Both the SST and the TRIPPEL spectrograph were designed for wavelengths up to 10830 \AA,
but until our campaign no attempt to obtain scientific data there had
been done. Our results show that the performance is within specifications. A photon
budget using nominal values for losses from all components in the optical train,
detector QE from the product sheet, and a gain measured from the actual camera specimen, 
agree within 10\% of the observed count rates (Kiselman, private communication).


\subsection{Reduction}\label{reduc}
The reduction process is described in detail in \citetads{2009A&A...507..417P} with some adaptations described in \citetads{2011A&A...535A..14K}. Here we summarize the reduction process briefly.

Flat field images are obtained while moving in circles over quiet sun disk-center, with the AO giving randomized voltages and thus obtaining a quiet sun average spectrum. 

The geometric corrections are estimated by placing a grid in front of the slit. The obtained image contains the spectral lines deformed by smile (i.e. geometrical distortions in the spectral direction), and the grid lines deformed by keystone (i.e. geometrical distortions in the spatial direction). By fitting both the spectral and the grid lines, we make a map of the geometrical distortions and apply the corrections to all spectra.

To obtain a wavelength calibration, the spectra are compared with the FTS atlas (\citeads{1984SoPh...90..205N}). 
Using telluric lines in the spectral field-of-view, we estimate the error on this calibration to be smaller than 0.45 m s$^{-1}$, similar to what was found by \citetads{2009A&A...507..417P}. 

A stray-light correction is obtained by convolving a Gaussian instrumental profile of TRIPPEL with the FTS atlas. Comparing this with the flat-field spectra, we find an estimate for both the spectral resolution of TRIPPEL and the amount of stray-light. We have found that for H$\beta$ and He \textsc{i} 10830 \AA, the fitting algorithm usually converges to a spectral resolution of $R\sim 150\,000$ and an amount of stray-light around 5\%. However for He \textsc{i} D\textsubscript{3}, the procedure is not well constrained because of the lack of the strong lines in the spectral region. Therefore, no stray-light correction was applied to the He \textsc{i} D\textsubscript{3} spectra. The estimate for the spectral resolution has been used to rebin all spectra according to the Nyquist sampling theorem, as shown in Table~\ref{instr}.

\subsection{Telluric Correction \label{telcorrec}}
\begin{table}
\centering
\footnotesize
\caption{Spectral lines present in the wavelength region shown in Fig.~\ref{tellurics}. Helium line information is taken from the online NIST Atomic Spectra Database \citep{NIST_ASD}. All other lines are taken from \citet{moore}. \label{lines}}
\begin{tabular}{c c c }
\hline \hline
 & &    \\[-2mm]
Wavelength & Atom & Levels  \\[1mm]
in Air [\AA] & &  \\& &  \\[-2mm]\hline
 & &  \\[-2mm]
5875.143 & Atm -- &   \\[1mm]
5875.444 & Atm H\textsubscript{2}O &   \\[1mm]
5875.596 & Atm H\textsubscript{2}O &  \\[1mm]
5875.599 & He \textsc{i} & 1s2p \textsuperscript{3}P\textsubscript{2} - 1s3d \textsuperscript{3}D\textsubscript{1}  \\[1mm]
5875.614 & He \textsc{i} & 1s2p \textsuperscript{3}P\textsubscript{2} - 1s3d \textsuperscript{3}D\textsubscript{2}  \\[1mm]
5875.615 & He \textsc{i} & 1s2p \textsuperscript{3}P\textsubscript{2} - 1s3d \textsuperscript{3}D\textsubscript{3}  \\[1mm]
5875.625 & He \textsc{i} & 1s2p \textsuperscript{3}P\textsubscript{1} - 1s3d \textsuperscript{3}D\textsubscript{1}  \\[1mm]
5875.640 & He \textsc{i} & 1s2p \textsuperscript{3}P\textsubscript{1} - 1s3d \textsuperscript{3}D\textsubscript{2}  \\[1mm]
5875.966 & He \textsc{i} & 1s2p \textsuperscript{3}P\textsubscript{0} - 1s3d \textsuperscript{3}D\textsubscript{1}  \\[1mm]
5876.124 & Atm H\textsubscript{2}O &   \\[1mm]
5876.296 & Fe \textsc{i} &  \\[1mm]
5876.449 & Atm H\textsubscript{2}O &  \\[1mm]
5876.556 & Cr \textsc{i} &  \\
& &  \\[-2mm]
\hline
\end{tabular}
\end{table}

\begin{figure}
\includegraphics[scale=1]{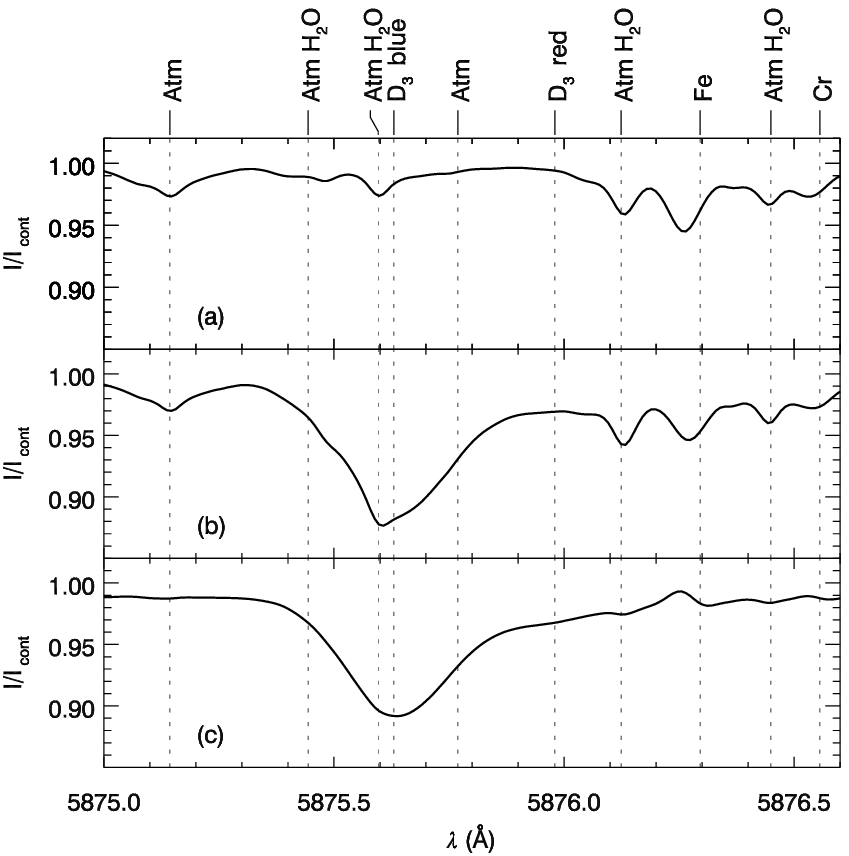}
\caption{Demonstration of the correction for telluric blends with the He \textsc{i} D\textsubscript{3} line. Panel \textbf{(a)} shows the reference quiet sun spectrum containing the blends but not any He \textsc{i} D\textsubscript{3} signal. Panel \textbf{(b)} shows the original spectral region including blends in He \textsc{i} D\textsubscript{3}. Panel \textbf{(c)} shows the corrected He \textsc{i} D\textsubscript{3} line (including the artefact at 5876.3 \AA\;resulting from telluric correction). \label{tellurics}}
\end{figure}

The He \textsc{i} D\textsubscript{3} line consist of six components. Five of those are blended together at $\lambda=5875.562$ \AA. The sixth component is shifted to the red at $\lambda=5875.913$ \AA, see Table~\ref{lines}. The line is contaminated by several blends: a telluric H\textsubscript{2}O blend right on top of the line core and several telluric and solar blends in both wings (see Fig.~\ref{tellurics} panel (b), and Table~\ref{lines}). During the observations, we measured telluric lines by taking quiet sun reference spectra, which are entirely void of He \textsc{i} D\textsubscript{3} signal, as shown in Fig.~\ref{tellurics}, panel (a). However, due to differences in airmass, the strength of the telluric lines changes too fast with time to use these reference spectra directly to correct for the telluric blends in all obtained raster scans.

To separate between telluric signal and He \textsc{i} D\textsubscript{3} signal, we apply the technique of Principal Component Analysis (PCA). When decomposing reference spectra, the first principal component represents the average position, width and strength of the spectral lines in the spectral region, except the He \textsc{i} D\textsubscript{3} line which is not present in these quiet sun reference spectra. Afterwards, the science data of one raster scan is projected on this first principal component. The projection correctly represents the position, width and strength of the telluric lines in the used raster scan, without containing any He \textsc{i} D\textsubscript{3} signal. Therefore, the projection is used to correct the raster scan for telluric blends. The result is shown in Fig.~\ref{tellurics}, panel (a). This technique works well to correct for telluric blends, but it badly affects the solar lines in the spectral region, see the artefact at 5876.3 \AA\, in Fig.~\ref{tellurics}, panel (c).

\subsection{Alignment and Co-Observations}\label{align}
With the SST, we acquired 10 consecutive raster scans in H$\beta$, He \textsc{i} D\textsubscript{3} and He \textsc{i} 10830 \AA. After completing a scan, the correlation tracker could not always perfectly recover the initial raster scan position. Therefore, the different scans needed alignment in time. We also took into account the effect of beam rotation in the telescope. We aligned the different spectral regions regarding their different FOV and corrected for differential atmospheric refraction.

The co-observations with IRIS are obtained with a large dense 96-step raster with a FOV of 33$\times$119\arcsec\,, observing the Mg \textsc{ii}, Si \textsc{iv}, O \textsc{i} and C \textsc{ii} spectral regions. The cadence of the IRIS raster scan is 534s. Using cross-correlations, we aligned the continua of the IRIS rasters with the continua of the SST rasters.

Data from SDO/AIA and from the Helioseismic and Magnetic Imager (HMI, \citeads{2012SoPh..275..229S}) on board of SDO was aligned with the SST data using the HMI continuum intensity image.


\section{Data Analysis}\label{sec:ana}
\subsection{Ellerman bomb Detection and Light Curves}\label{subs:det}
\begin{figure}
\includegraphics[scale=1]{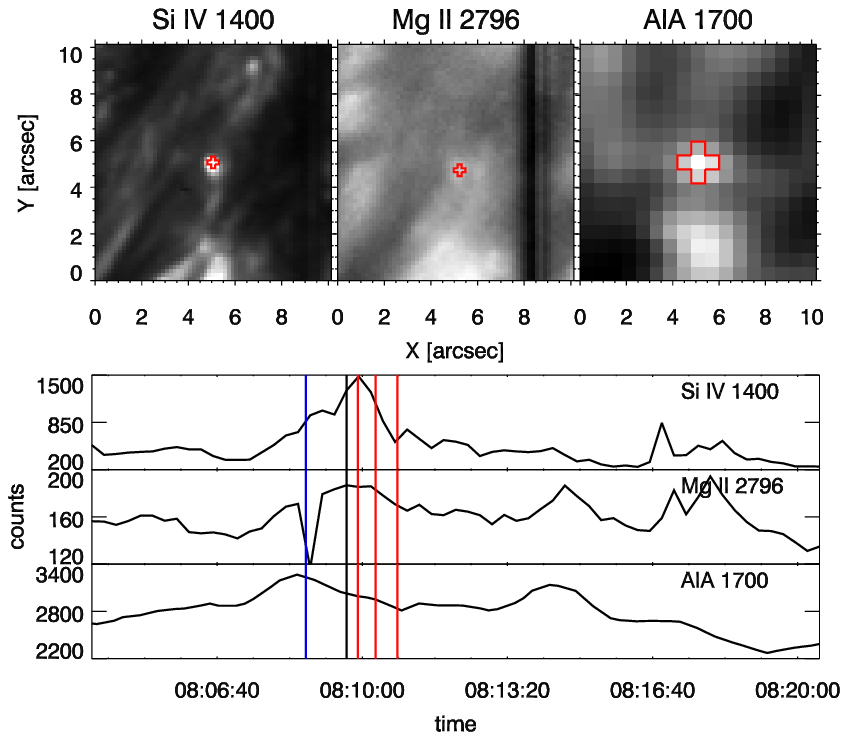}
\caption{Light curves for EB1. The upper three panels show IRIS and SDO/AIA images of the EB. The red cross indicates the pixels used to calculate the corresponding light curves shown in the lower three panels. The vertical blue line overplotted on the light curve indicates the time where the IRIS slit was scanning the EB. The vertical black line indicates the time at which the images in the upper three panels are shown. The three vertical red lines show the times of the SST slit scanning the EB in H$\beta$ (left), He \textsc{i} D\textsubscript{3} (middle) and He \textsc{i} 10830 \AA (right). The time difference is due to differential atmospheric refraction. A movie of this figure is available. \label{LC1}}
\end{figure}
\begin{figure}
\includegraphics[scale=1]{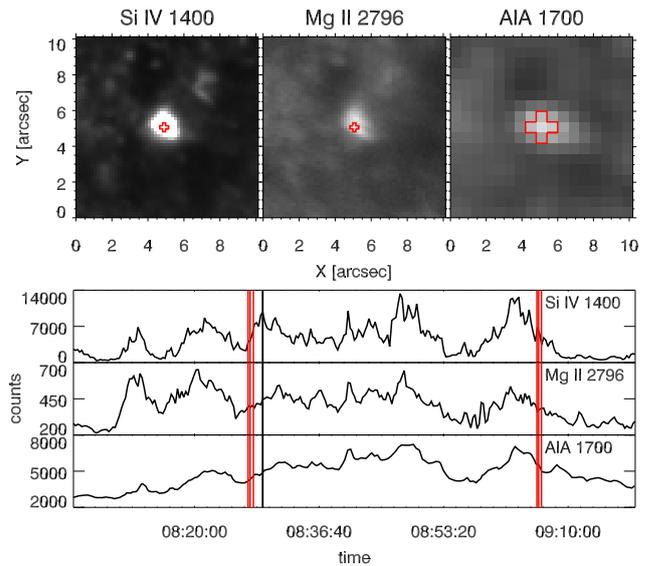}
\caption{Same as Fig.~\ref{LC1} but for EB2. EB2 was scanned twice by the SST slit, therefore in our further analysis, we refer to EB2a for the first event and EB2b for the second event. EB2 was not scanned by the IRIS slit. Movie available.\label{LC2}}
\end{figure}
\begin{figure}
\includegraphics[scale=1]{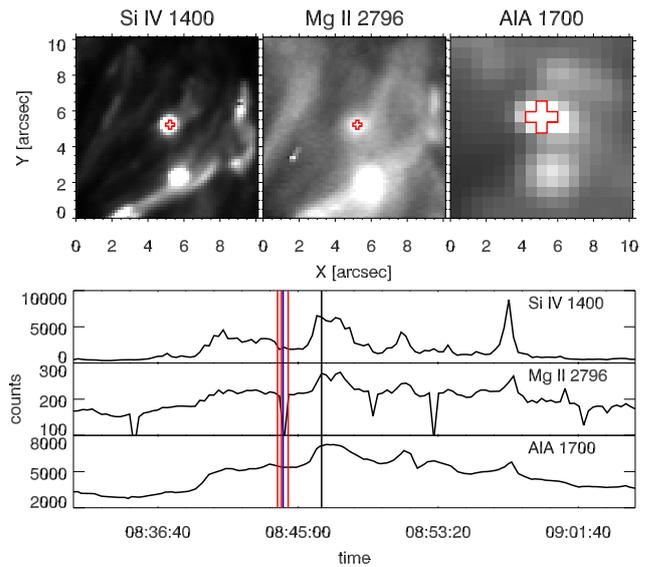}
\caption{Same as Fig.~\ref{LC1} but for EB3. Movie available.\label{LC3}}
\end{figure}

We use a semi-automated method to detect all EB candidates in our data based on properties of the H$\beta$ profile.
\begin{enumerate}
\item All H$\beta$ profiles are normalized to their local continuum. Then we have constructed an image consisting of the average of the blue wing of H$\beta$ between H$\beta-1.1$\AA\,and H$\beta-0.55$\AA\,and the red wing between H$\beta+0.55$\AA\,and H$\beta+1.1$\AA. In this wing image, we select all the pixels with a brightness exceeding 30\% over the local continuum.
\item The selected pixels are grouped together by combining adjoining pixels into events. For each event, we selected the pixel with maximum wing brightness as the location of the EB candidate.
\item For each EB candiate, the H$\beta$ line core intensity is compared to the line core intensity of an average H$\beta$ profile in the quiet sun. For this comparison, neither the event profile nor the quiet sun profile are divided by their local continuum. We selected only those events which show less than 25\% difference between EB candidate and quiet sun.
\end{enumerate}

Finally, we ended up with 21 identified EBs in the 10 raster scans, which are shown with plus symbols in Fig.\,\ref{panels}. Some of those 21 EBs are part of the same event observed in the several raster scans. Since EBs show complex behaviour in time such as disappearing and reappearing, splitting and migrating, we decided to show all events at each time step instead of relating one time step to another, which is hard using raster scans with 11min cadence. The only way to obtain more insight in time evolution of the EBs is to use higher cadence data.

In this paper, we focus on four EBs which show emission signals in He \textsc{i} D\textsubscript{3}. The other 17 EBs in our data show no evidence of emission signals in helium. In Sec.~\ref{subs:cond}, we discuss possible reasons of why those 17 EBs have no emission in their helium spectra.

The four EBs which show emission signals are indicated in Fig.\,\ref{panels} with a yellow plus: EB1 at $x=(15,21)\arcsec$\,, EB2a at $x=(31,28)\arcsec$\,, EB2b at $x=(30,27)\arcsec$\, and EB3 at $x=(14,18)\arcsec$\,. EB2a and EB2b correspond to the same EB at different time steps. For all these bombs, we have computed light curves using IRIS slit-jaw movies in Si \textsc{iv} 1400 \AA\, and Mg \textsc{ii} 2796 \AA, and the SDO/AIA 1700 \AA\,channel. The light curves are shown in Figs.\,\ref{LC1}, \ref{LC2} and \ref{LC3}. 

To construct the light curves, we used non-rotated IRIS and SDO images with their original pixel scale. Therefore, the alignment between the IRIS slit-jaw images and the SDO/AIA image shown in Figs.\,\ref{LC1}, \ref{LC2} and \ref{LC3} is not exact. The value for the light curve at each timestep is calculated as the mean of the 5 pixels indicated with a red cross in Figs.~\ref{LC1}, \ref{LC2} and \ref{LC3}, of which the central pixel is the brightest pixel in a limited area around the EBs. This method results in an intensity count which is independent of the area of the EB. We obtain qualitatively similar light curves for all channels. The strong signal in SDO/AIA 1700 \AA\ is further proof that we have detected true EBs (\citeads{2013ApJ...774...32V}). The three EBs are distinctly visible in the IRIS Si \textsc{iv} slit-jaw images, and visible, however more diffuse, in the Mg \textsc{ii} slit-jaw images.

The exact times at which the SST slit is scanning the bomb, are indicated with red vertical lines in Figs.\,\ref{LC1}, \ref{LC2} and \ref{LC3}. We always scan first H$\beta$ (left), then He \textsc{i} D\textsubscript{3} (middle) and He \textsc{i} 10830~\AA\;(right) last. The time difference between scanning times of different spectral lines is due to differential atmospheric refraction. The blue line corresponds to the IRIS slit scanning the bomb. All EBs were captured by IRIS slit-jaw images, but only EB1 and EB3 were covered by the IRIS slit.

The light curves and movies show different time evolution and properties for the three EBs. EB1 is short lived and has a small size of $\la$ 1\arcsec\, in diameter. EB1 is a single event of appearing, brightening and disappearing of the bomb with a lifetime of the order of 10 min. EB2 on the other hand is long-lived with a lifetime of over one hour and has a larger size of around 2\arcsec . The movies reveal that EB2 has a complex morphology and time evolution in which it appears, migrates, disappears, reappears and at some point splits into two bombs\footnote{Given the complex evolution of EB2, the question whether we are still observing the same EB2 after one hour becomes a more philosophical one (similar problem as the river of Heraclitus).}. If EB1 and EB2 are on opposite extremes in terms of life time, evolution and morphology, then EB3 is ranged in the middle between those two. It has a size of around 1\arcsec\, in diameter and a life time of around 30 min. EB3 shows several maxima in intensity but never fully disappears in between those maxima.

Given the relatively large size and the long lifetime of EB2, the question might be raised whether this is a true EB. The detection of strong co-spatial signal in SDO/AIA 1700 \AA\;but not in any other AIA channels, argues in favour of a true EB detection.

\citeads{2015ApJ...812...11V} have also defined an event called ``flaring arc filament'' (FAF). These events are slightly more elongated and shorter lived, and their Balmer line spectra show intensity enhancement in the absorption core. FAFs could hence be interpreted as high-energy EBs breaking through the chromospheric canopy of fibrils. We do not classify EB2 as a FAF because we have not detected intensity enhancements in the line core of H$\beta$. Also, according to \citeads{2015ApJ...812...11V}, FAFs are bright in SDO/AIA 1600 \AA\;but less evident in SDO/AIA 1700 \AA, which is not the case for EB2.

We should keep in mind that the limited spatial resolution of SDO/AIA makes the EBs appear larger than they actually are. This is clear when we compare to higher resolution IRIS slit-jaw images in which the events look smaller than in the SDO/AIA 1700 \AA\;channel. Similar arguments hold for the long lifetime of EB2. The temporal cadence of the SDO/AIA images (24s) and the IRIS slit-jaw images (17s) will not completely resolve the event into several shorter lived events at the same location. However, it is clear from the movie that EB2 is a combination of reoccurring events at the same spatial position, which explains the longer than usual lifetime of EB2.

\subsection{Ellerman bomb spectra \label{subs:spec}}

\begin{figure*}
\includegraphics[scale=1]{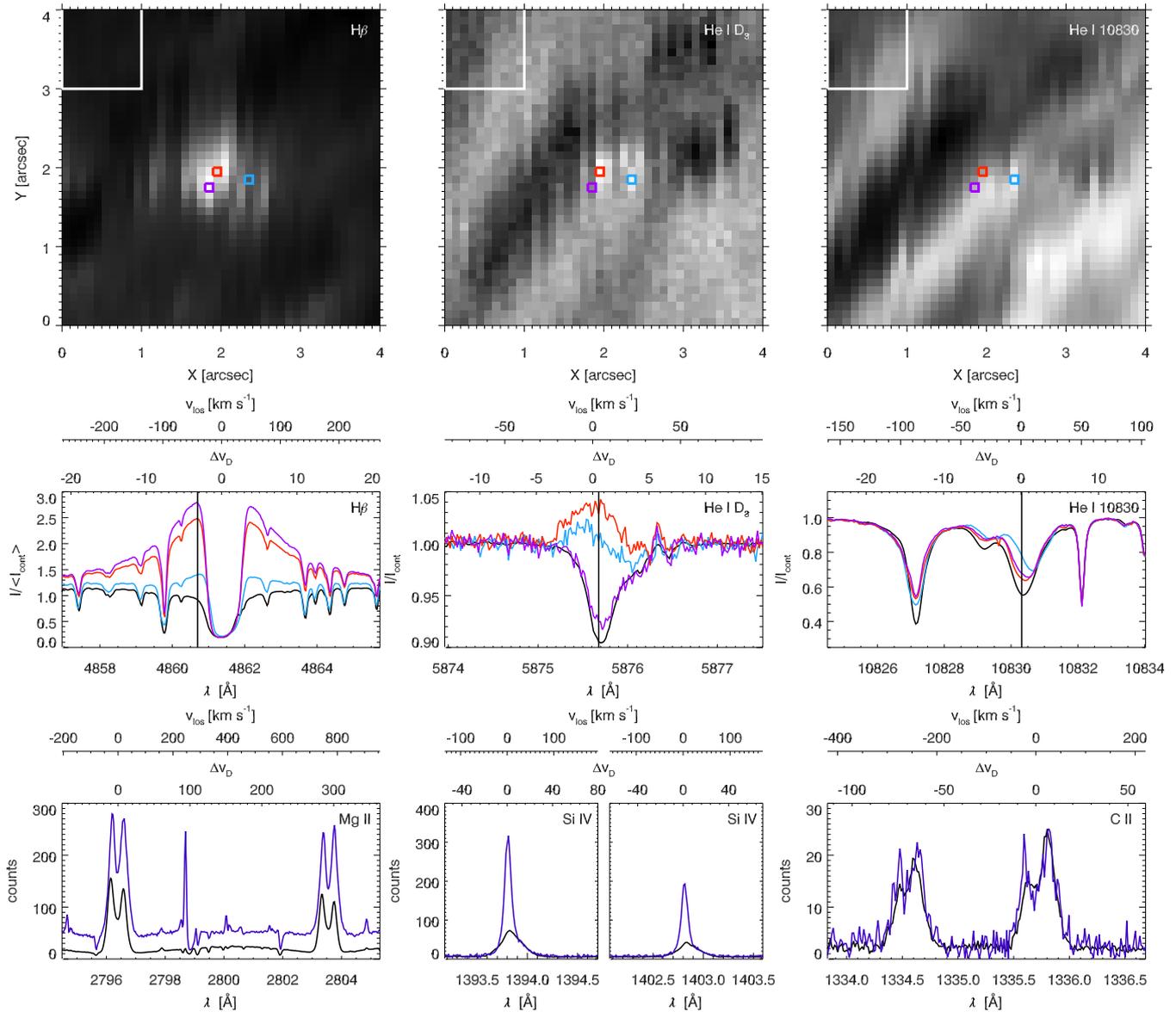}
\caption{Spectra and raster scan images of EB1. The upper three panels represent raster scan images for respectively H$\beta$, He \textsc{i} D\textsubscript{3}, and He \textsc{i} 10830 \AA. The exact wavelength at which the raster scan images are shown is indicated by a black vertical line on the corresponding spectra. The white box indicates the region for which the average spectra is given in black. The colored boxes around certain pixels correspond to the pixels for which the spectra are given in the same color. All spectra are given on three scales: wavelength $\lambda$, Doppler width $\Delta v_{\rm D}$ and line-of-sight velocity $v_{\rm LOS}$. \label{spec_withim}}
\end{figure*}
\begin{figure*}
\includegraphics[scale=1]{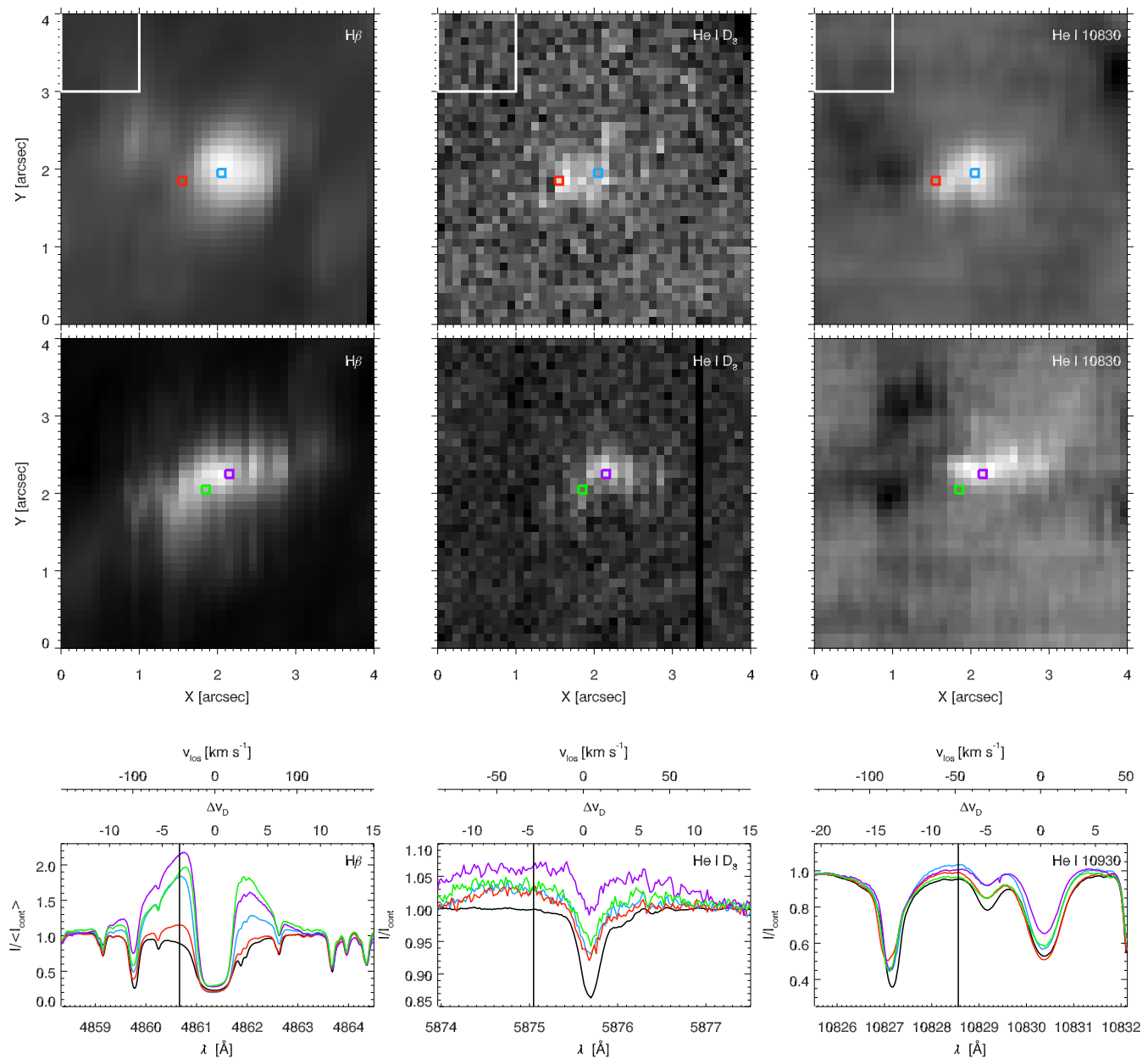}
\caption{Same as Fig.~\ref{spec_withim} but for EB2. EB2 was not captured by the IRIS slit. The upper three raster scan images correspond to EB2a and the lower three to EB2b. \label{spec_withim5}}
\end{figure*}
\begin{figure*}
\includegraphics[scale=1]{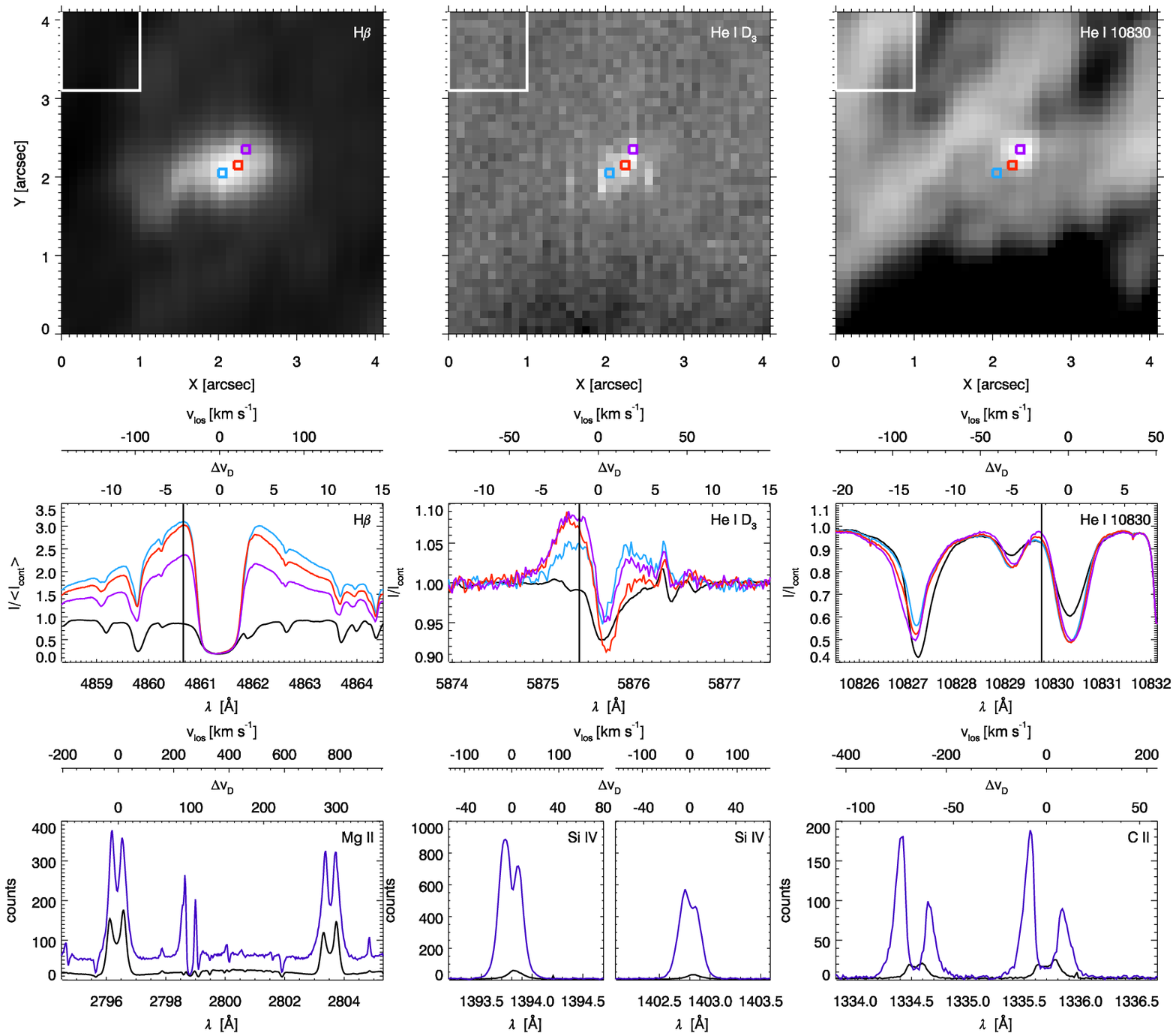}
\caption{Same as Fig.~\ref{spec_withim} but for EB3. \label{spec_withim10}}
\end{figure*}
We show raster scan images and spectra of EB1, EB2 and EB3 in Fig.\,\ref{spec_withim}, Fig.\,\ref{spec_withim5} and Fig.\,\ref{spec_withim10} respectivily. All spectra are shown on a wavelength, los-velocity ($v_{\rm LOS}$) and Doppler width ($\Delta  \nu_D$) scale, with \begin{equation}
\Delta \nu_D=\frac{\lambda-\lambda_0}{\lambda_0}\cdot\frac{c}{\sqrt{\frac{2kT}{m}}},
\end{equation}
with $\lambda_0$ the central wavelength, $c$ the speed of light, $k$ the Boltzmann constant, $m$ the mass of the atom and $T=10^4$\,K. The chosen value of $T$ is arbitrary, but does not affect the interpretation of the scale as long the same value is used everywhere and it is only used for relative comparison.

\paragraph{{He \textsc{i} D\textsubscript{3} spectra}} The He \textsc{i} D\textsubscript{3} spectra of EB1, EB2 and EB3 (Figs.\,\ref{spec_withim}, \ref{spec_withim5} and \ref{spec_withim10} respectivily) show profiles which are a combination of emission and absorption. The emission features for all EBs appear broadened and slightly blue-shifted, however the broadening and shift seems much larger in case of EB2. In EB1, the absorption and emission component appear blended on top of one another resulting in almost flat profiles or profiles with a $\sim$-shape, due to small velocity differences. The two components of EB2 are more easily spotted since the EB2 emission component is extremely broad while the absorption component has a more conventional shape comparable to regions outside the bomb. The same is true for EB3 where the absorption component seems unexceptional while the emission component is very broad. It is also remarkable that the spectra of EB2a and EB2b have a similar appearance, even though the time difference between them is around 30 min.

All He \textsc{i} D\textsubscript{3} spectra show an artefact at 5876.3 \AA. This is due to a solar blend with an iron line, wrongly corrected by our telluric correction, see Sec.~\ref{telcorrec}.

The emission component of the He \textsc{i} D\textsubscript{3} line must be linked to the EB itself which is expected to be located in the upper photosphere or lower chromosphere, while the absorption component probably originates in the upper chromosphere where He \textsc{i} D\textsubscript{3} is usually formed. This situation resembles Balmer line EB profiles, where the wings are in emission and associated with the EB, while the core is in absorption formed in higher layers. 

\paragraph{He \textsc{i} 10830~\AA\;spectra} He \textsc{i} 10830~\AA\;is the resonance line of the neutral helium triplet state between atomic energy levels 1s2s\textsuperscript{3}S and 1s2p\textsuperscript{3}P\textsuperscript{o}. Its upper level corresponds to the lower level of He \textsc{i} D\textsubscript{3}, which is the transition between 1s2p\textsuperscript{3}P\textsuperscript{o} and 1s3d\textsuperscript{3}D. Hence the He \textsc{i} 10830~\AA\; line is stronger but we expect the lines to behave similarly. 

However, the emission features in He \textsc{i} 10830~\AA\,are much harder to detect since the wings of the line are blended with the strong Si line on the blue side and a strong telluric on the red side. In EB1, we do not observe any emission signature in He \textsc{i} 10830~\AA\,at all, only a strong asymmetry in the absorption profile. In some of the pixels, there seems to be a hint of emission, but the profiles never exceed the continuum level. Looking at the light curve of EB1, we see that He \textsc{i} 10830~\AA\,has been recorded approximately 30 seconds after He \textsc{i} D\textsubscript{3}, when the EB event is weaker, which might explain the absence of emission. In EB2, we detect emission signals in some pixels in the blue wing of He \textsc{i} 10830 \AA, with a hint of emission in the red wing as well. There is however always a strong absorption component present in the line core. In EB3, the emission signal is not exceeding the continuum level, but the profiles have some properties indicating that an emission component is affecting the line. The latter can be seen clearly in the case of EB3 in Fig.~\ref{spec_withim10}, upper right panel. The raster scan close to the blue continuum shows the bomb as a brighter feature compared to its surroundings at 10830 \AA.

We conclude that an emission component is present in He \textsc{i} 10830~\AA\;but it is hard to detect it, due to blends with strong lines in both wings, and a strong absorption component caused by upper-chromospheric material.

\paragraph{H$\beta$ spectra} The H$\beta$ profiles of EB1, EB2 and EB3 indicate that the line core is unaffected by the bomb, when we compare with an average H$\beta$-profile in the immediate surroundings of the bomb. EB1 and EB3 have symmetrical H$\beta$-profiles while EB2a shows a blue asymmetry. Asymmetries have been reported in H$\alpha$ (by e.g. \citeads{1972SoPh...26...94B,1983SoPh...87..135K,1997A&A...322..653D}), with blue asymmetries prevailing. The asymmetries are supposedly caused by velocity shifts in the fibrils, due to the inverse Evershed effect (\citeads{1972SoPh...26...94B,2013JPhCS.440a2007R}). However, this effect should only play a role in superpenumbral areas and it does not explain the dominance of blue asymmetries. \citetads{1983SoPh...87..135K} and \citetads{1997A&A...322..653D} also reported cases where the emission component associated with the EB itself is blueshifted. We think the latter case is more likely to explain the observed blue asymmetry in EB2, since EB2 is located outside the superpenumbral area, see Fig.~\ref{panels}, panel a.

\paragraph{Raster Scan Images} The raster scan images of the bombs in He \textsc{i} D\textsubscript{3} show that we are able to observe the bombs as a round emission feature, if we select the right wavelength position. The same seems to be true for He \textsc{i} 10830~\AA, but in EB1 the bomb is mostly invisible. We have selected the wavelength positions to show the raster scan images at large contrast, which is different for the three bombs due to different Doppler shifts of the emission component.

The area of the bombs as observed in the helium triplet lines is smaller than in H$\beta$ as seen in Figs.~\ref{spec_withim}, \ref{spec_withim5} and \ref{spec_withim10}. The EBs in helium lines have lower contrast than in H$\beta$. These two reasons probably explain why helium emission signals have not been observed before in helium lines. Very high spatial resolution and good seeing conditions appear to be needed to detect EBs in helium lines. 

\paragraph{IRIS spectra} EB1 and EB3 are covered by the IRIS slit. The IRIS spectra of EB1 are shown in Fig.~\ref{spec_withim}. Emission in the subordinate triplet Mg \textsc{ii} 2799 \AA\;is strongly enhanced, indicating the presence of a steep temperature rise (at least 1500 K) in the lower chromosphere (\citeads{2015ApJ...806...14P}). We also observe a redshift of the core of this line, resulting in a strong asymmetric profile. The Mg \textsc{ii} h\&k lines are brighter and slightly broader compared to profiles in the surroundings of the bomb. The Si \textsc{iv} lines are strongly enhanced but not very broad and we do not observe double peaked profiles. The ratio at maximum intensity of the two Si \textsc{iv} lines is $\sim 1.6$, hence the line is not completely optically thin (the optically thin ratio equals two). The C \textsc{ii} profiles are not clearly affected by EB1 and do not differ substantially from profiles in the immediate surroundings of EB1.

The IRIS profiles of EB3 are slightly more spectacular, especially the Si \textsc{iv} and C \textsc{ii} lines. The Si \textsc{iv} lines appear bright, double-peaked, broadened and slightly blue-shifted. The ratio of the peak intensity of the two Si \textsc{iv} lines is 1.5, suggesting a departure from the optically thin regime. The C \textsc{ii} lines are intense and strongly broadened. The triplet Mg \textsc{ii} 2799 \AA\; line shows strong emission again, looking more symmetrical than in EB1. The Mg \textsc{ii} h\&k lines are strong and have broadened wings.

\paragraph{SDO/AIA images} We have aligned and compared all AIA channels with our EBs and have found no signature of the bombs in any other channels than AIA 1700 \AA\, and AIA 1600 \AA. 

\subsection{Inversion of the He emission profiles \label{subs:inv}}
Estimating physical parameters of an EB atmosphere requires fitting of the profiles. We focus on the He \textsc{i} D\textsubscript{3} profiles, since the emission component is more pronounced than in the He \textsc{i} 10830~\AA\;profiles. Two inversion codes are currently available to model helium triples lines: \textsc{HeLIx}\textsuperscript{+} (\citeads{2004A&A...414.1109L}) and \textsc{Hazel} (\citeads{2008ApJ...683..542A}). 

Our observations have clearly shown profiles which are a combination of emission and absorption. \textsc{HeLIx}\textsuperscript{+} allows fitting multiple components assuming several Milne-Eddington atmospheres that are combined using a filling factor. For two components, the emergent intensity $I_\nu$ is given by
\begin{dmath}\label{general}
I_{\nu}=\alpha\cdot\int_0^{\infty}S_{\nu ,1}(t_{\nu})e\,^{-t_\nu}\,{\rm d}t_\nu
 + (1-\alpha)\cdot\int_0^{\infty}S_{\nu ,2}(t_{\nu})e^{-t_\nu}{\rm d}t_\nu,
\end{dmath}
with $\alpha$ the filling factor, $S_\nu$ the source function and $t_\nu$ the integration variable corresponding to optical depth. Indices 1 and 2 correspond to the two components (emission and absorption) in the atmosphere. 

The \textsc{Hazel} code assumes constant property slabs, each having a different but constant value for the source function. The absorption and emission component can be combined with a filling factor. The emergent intensity is then given by
\begin{dmath}\label{hafill}
I_\nu=\alpha\cdot\left[I_{\nu,c} e^{-\tau_{\nu,1}}+S_{\nu,1}\left(1-e^{-\tau_{\nu,1}}\right)\right]+(1-\alpha)\cdot\left[I_{\nu,c}e^{-\tau_{\nu,2}}+S_{\nu,2}\left(1-e^{-\tau_{\nu,2}}\right)\right].
\end{dmath}
The continuum intensity $I_{\nu,c}$ acts as a boundary condition and depends on the viewing angle.

\textsc{Hazel} also allows one to stack the two slabs, one on top of the other, and use the emergent intensity $I_\nu(\tau_{\nu,1})$ of the first slab as a boundary condition for the second slab. The emergent intensity for this case equals
\begin{dmath}\label{hastack}
I_\nu= I_{\nu,c}e^{-(\tau_{\nu,1}+\tau_{\nu,2})}+S_{\nu,1}\left(1-e^{-\tau_{\nu,1}}\right)e^{-\tau_{\nu,2}}+S_{\nu,2}\left(1-e^{-\tau_{\nu,2}}\right).
\end{dmath}
This equation has also been used by \citet{2014ApJ...792...13H} in their two-cloud model to fit H$\alpha$ EB profiles. Their model uses the same idea: emission in the line wings is modelled with a lower cloud and absorption in the line core with an upper cloud.

Combining components using a filling factor such as in Eqs.~\ref{general} and \ref{hafill}, accounts for stray-light and seeing effects where light from one pixel gets smeared out over surrounding pixels, or when the EB is not fully resolved in a pixel. When using two stacked slabs, as in Eq.~\ref{hastack}, it is assumed that the emission from the EB originates in a deeper layer (in this case the upper photosphere or lower chromosphere) and then penetrates a higher layer (the upper chromosphere) where absorption affects the profile. Eqs.~\ref{hafill} and \ref{hastack} show that stacking slabs or using a filling factor are physically and analytically different. The question is whether these different approaches yield compatible results for the physical parameters in which we are interested, such as the Doppler motions $v_{\rm Dop}$ and the LOS velocity $v_{\rm LOS}$. 

\begin{figure*}
\includegraphics[scale=1]{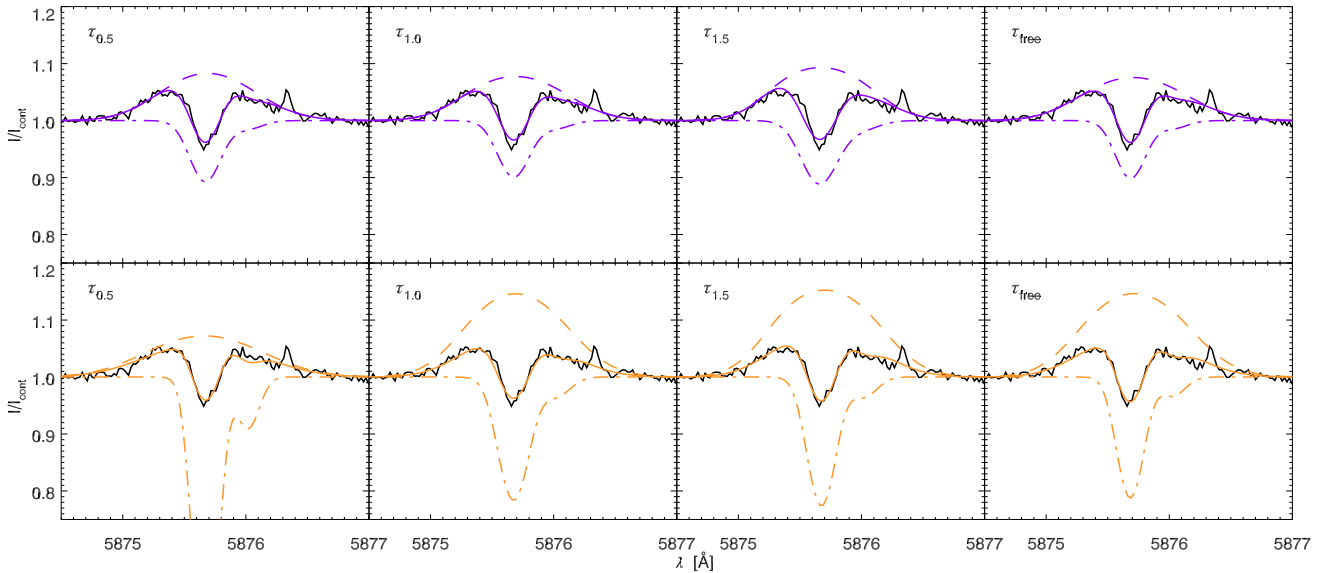}
\caption{In every panel, we show the same observed EB3 profile in a black solid line. Upper row (purple): fits obtained using two stacked slabs. Lower row (orange): fits obtained using a filling factor. The total fit is shown in color in a solid line. Then the emission component (dashed line) and the absorption component (dash-dot line) are overplotted. The $\tau$ label indicates whether $\tau_1$ has been fixed to a certain value, or it has been free to vary in the inversion, see Table~\ref{tabelslabfill}. \label{slabfill}}
\end{figure*}

We proceed using \textsc{Hazel}, since the code supports both filling factors and stacking slabs, and study the difference in the resulting profiles. In Fig.~\ref{slabfill}, we show fits for a series of fixed values of $\tau$ to the same observed EB profile. The fits obtained using two stacked slabs generally consist of a weaker absorption and emission component than the ones combined with a filling factor. This is to be expected, since the total emergent profile is not an average of both components. Good fits could be obtained for different values of $\tau$, both with using a filling factor, as well as using stacked slabs, hence degeneracy is present. 

In the end, we chose to fit our EB profiles using two stacked slabs. We expect this approach to represent the physics better: we believe that the absorption component gets imprinted on the emission profile in higher layers and this effect is probably dominating over seeing and resolution effects. Also, the absorption component in the stacked slab case appears more similar to the absorption profiles outside the EB region. Another advantage of the two stacked slab approach is that it does not have the filling factor as an extra free parameter. 

\begin{table}
\centering
\begin{scriptsize}
\caption{Retrieved parameters for the two component inversion in Fig.~\ref{slabfill}\label{tabelslabfill}. The upper list of parameters in the table corresponds to the purple fits in Fig.~\ref{slabfill}\label{tabelslabfill} using two stacked slabs and the lower list of parameters corresponds to the orange fits using a filling factor $\alpha$. The parameters in bold have been held constant during the inversion.}
\begin{tabular}{l r r r r c}
\hline\hline
  & & & & & \\[-2mm]
  & $\tau_{0.5}$ & $\tau_{1.0}$ & $\tau_{1.5}$ & $\tau_{\rm free}$ & ranges \\[1mm]\hline
  & & & & & \\[-1mm]
$\tau_1$            & \textbf{0.50}   & \textbf{1.00}  & \textbf{1.50} &  0.81 & 0 -- 2 \\[1mm]
${v_{\rm Dop,1}}\tablefootmark{1}$ & 24.13  & 23.19 & 21.07 &  24.76 & 10 -- 40 \\[1mm]
${v_{\rm LOS,1}}^1$ & 3.23   & 3.38  & 2.49  &  4.26  & -17.5 -- 22.5 \\[1mm]
$\beta_1$           & 3.62   & 3.36  & 3.35  &  3.39  & 3.0 -- 3.7  \\[1mm]
$\tau_2$            & 0.18   & 1.02  & 0.17  &  0.17  & 0 -- 1 \\[1mm]
${v_{\rm  Dop,2}}^1$ & 10.67  & 8.83  & 10.20 &  8.34  & 5 -- 11 \\[1mm]
${v_{\rm LOS,2}}^1$ & 3.62   & 2.84  & 3.01  &  4.37  & -17.5 -- 22.5\\[1mm]
$\beta_2$           & 0.98   & 1.02  & 0.79  &  0.97  & 0 -- 2 \\[1mm]   \hline
  & & & & & \\[-1mm]
$\tau_1$            & \textbf{0.50}   & \textbf{1.00}  & \textbf{1.50} &  1.15 & 0 -- 2 \\[1mm]
${v_{\rm Dop,1}}^1$ & 29.82  & 23.61 & 22.75 &  24.17 & 10 -- 40 \\[1mm]
${v_{\rm LOS,1}}^1$ & 1.72   & 3.42  & 4.05  &  4.72  & -17.5 -- 22.5 \\[1mm]
$\beta_1$           & 3.54   & 3.68  & 3.57  &  3.63  & 3.0 -- 3.7  \\[1mm]
$\tau_2$            & 0.86   & 1.53  & 0.48  &  0.52  & 0 -- 1 \\[1mm]
${v_{\rm Dop,2}}^1$ & 6.33   & 8.13  & 7.56  &  7.12  & 5 -- 11 \\[1mm]
${v_{\rm LOS,2}}^1$ & 3.76   & 3.82  & 3.85  &  4.29  & -17.5 -- 22.5\\[1mm]
$\beta_2$           & 0.32   & 1.40  & 1.20  &  1.39  & 0 -- 2\\[1mm]
$\alpha$            & 0.81   & 0.49  & 0.49  &  0.47  & 0 -- 1 \\[1mm]   \hline
\end{tabular}
\newline
\tablefoottext{1}{Velocities are given in km~s$^{-1}$. Positive LOS velocities correspond to redshifts.}
\end{scriptsize}
\end{table}

When trying to fit absorption and emission simultaneously, nearly any amount of emission can be compensated with any amount of absorption, leading to a wide range of non-unique solutions. It is obvious that the problem would be much better constrained by the observation of all Stokes parameters. In an attempt to limit the degeneracy, we try to fit the profiles with the emission component as small as needed to reproduce the profiles. This results in setting the source function to an upper limit of around 3.7 $I_{\nu,\rm c}$ or less. We have fixed the damping to $a=10^{-4}$, which is a typical value for the damping at the temperature minimum. The value of $a=10^{-4}$ for the damping parameter has also been used for He \textsc{i} 10830 \AA\; by \citet{2004A&A...414.1109L}, who investigated the influence of the damping on the fitting parameters for \textsc{HeLIx}\textsuperscript{+}.

In Fig.~\ref{slabfill} and Table~\ref{tabelslabfill}, we study the influence of different values of $\tau$ to the other fitting parameters. In the case of two stacked slabs (purple fits), the Doppler motions of the emission component $v_{\rm Dop,1}$ vary between 21 -- 25 km~s$^{-1}$ and the LOS velocity $v_{\rm LOS,1}$ between 2 -- 4 km~s$^{-1}$, with positive values corresponding to a redshift. It is clear that there is uncertainty in the retrieved parameters due to degeneracy, but we believe that we can obtain a reliable order of magnitude estimate using \textsc{Hazel}.

\begin{figure*}
\includegraphics[scale=1]{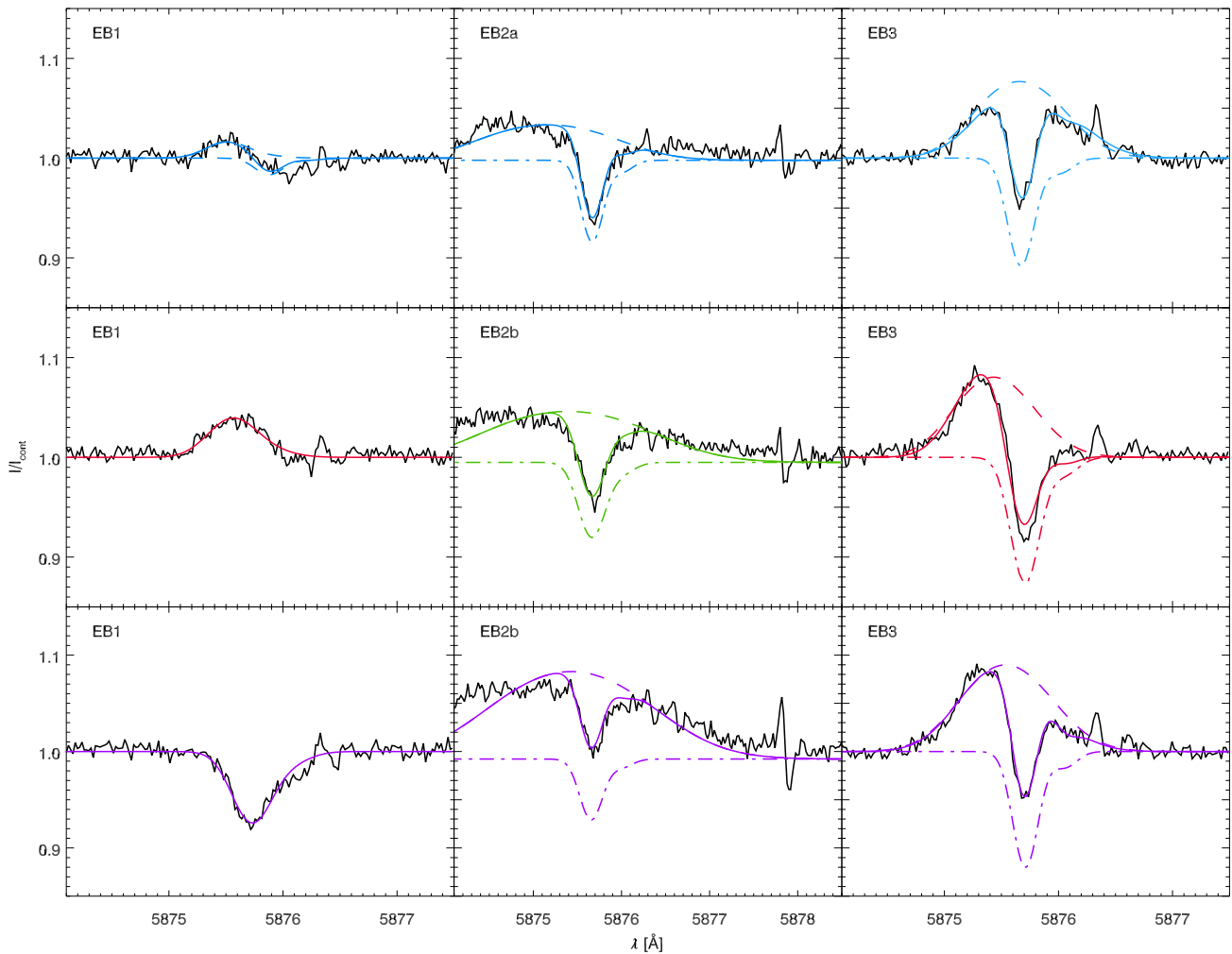}
\caption{Inversion results for EB profiles. The observed profile is shown as a black solid line. The colors of the fits correspond to the colors of the spectra, shown in Figs.~\ref{spec_withim}, \ref{spec_withim5} and \ref{spec_withim10}. The total fit is given as a colored solid line. The emission component (dashed line) and the absorption component (dash-dot line) are overplotted. The red and purple fit for EB1 have been obtained using only one component (solid line). The values for the parameters are shown in Table~\ref{tabelebfits}.\label{EBfits}}
\end{figure*}

The final results of the inversions with \textsc{Hazel} are shown in Fig.~\ref{EBfits} and Table~\ref{tabelebfits}. We have chosen to fit three EB profiles for each EB, corresponding to the profiles that we have shown in Fig.~\ref{spec_withim}, \ref{spec_withim5} and \ref{spec_withim10}. For the fitting results in Fig.~\ref{EBfits}, the observations are shown in black and the fitting results are shown in colors, corresponding to the colors of the profiles shown in Fig.~\ref{spec_withim}, \ref{spec_withim5} and \ref{spec_withim10}. We used \textsc{Hazel} with two stacked slabs  without magnetic field, hence we fitted 8 parameters of which 4 for each slab: optical depth $\tau$, Doppler motions $v_{\rm Dop}$, line-of-sight velocity $v_{\rm LOS}$ and the value of the constant source function $\beta$. 
Constant weights are used for the entire spectral region when the merit function $\chi^2$ is calculated. We found that the artefact at 5876.3 \AA\; did not influence our fitting results, hence we have not excluded it from the spectral range of the fit.

In Fig.~\ref{EBfits} for EB1, the blue profile is fitted with two components but since the emission and aborption component compensate one another to an almost flat observed profile, the fitting results are probably not reliable. The other two fits to EB1 shown in red and purple, could be fitted with one rather broad component only. The pure emission profile of EB1 (red), has a LOS velocity of ${v_{\rm LOS}}\sim -2.26$ km~s$^{-1}$ while the absorption profile (purple) is redshifted with ${v_{\rm LOS}} \sim 6.52$ km~s$^{-1}$.

The profiles of EB2 turn out to be badly fitted with constant slabs. The emission profiles are extremely broad, with ${v_{\rm Dop,1}}\sim 45$\, -- 56 km~s$^{-1}$, and there is an asymmetry present in the far blue at $\lambda=5874.0-5874.5$ \AA, possibly extending even further into the blue outside the observed spectral domain. The asymmetry cannot be fitted with a consant slab so it might be caused by velocity gradients in the emission component.

In EB3, we have obtained good fits with two constant slabs. The broadening of the emission component ranges between ${v_{\rm Dop,1}}=18$ -- 25 km~s$^{-1}$. In all EBs, it turns out that the emission component is blue-shifted with respect to the absorption component, except in one profile of EB3. There, only the fitted profile shown in blue color has a ${v_{\rm LOS}}\sim$ 4 -- 5 km~s$^{-1}$ for both the emission and absorption component.

\begin{figure}
\includegraphics[scale=1]{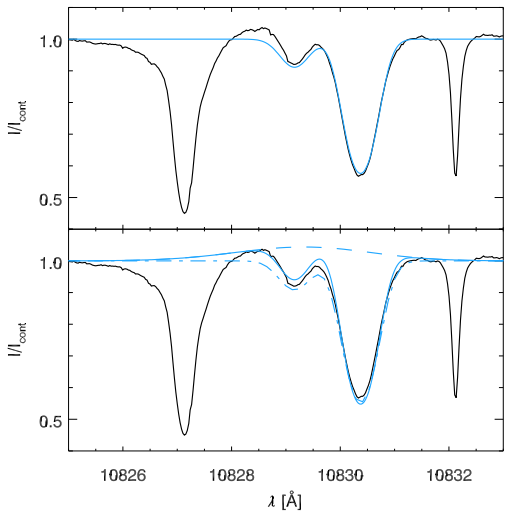}
\caption{Upper panel: 1 component inversion to a He \textsc{i} 10830~\AA\,EB2 profile, shown in blue in Fig.~\ref{spec_withim5}. Lower panel: 2 component inversion to the same profile. The total fit is given as a colored solid line. The emission component (dashed line) and the absorption component (dash-dot line) are overplotted. The ${v_{\rm LOS}}$ and ${v_{\rm Dop}}$ of the upper fit have been used to fix the values of the absorption component to the lower fit. The emission component values for ${v_{\rm LOS}}$ and ${v_{\rm Dop}}$ are fixed by the results for the He \textsc{i} D\textsubscript{3} fit. The values for the parameters are shown in Table~\ref{tabelsynth_10830}. \label{synth_10830}}
\end{figure}

So far, we have concentrated on the He \textsc{i} D\textsubscript{3} profiles since the emission is distinctly present. For He \textsc{i} 10830 \AA, the profiles are influenced by the emission component but the signal is mostly obscured by a strong absorption component and blends in both wings. In Fig.~\ref{synth_10830}, we show a He \textsc{i} 10830~\AA\; profile in which the emission is clearly present. We show that the profile is badly fitted with a one-component absorption profile. However, disentangling the two components with \textsc{Hazel} is difficult, since the extend of the overlap between the emission and the absorption component is not obvious. Therefore, we have assumed that the emission component has the same Doppler broadening ${v_{\rm Dop}}$ and the same $\Delta v_{\rm LOS}$ as the corresponding He \textsc{i} D\textsubscript{3} profile, with $\Delta v_{\rm LOS}=v_{\rm LOS,2}-v_{\rm LOS,1}$. The two-component fit shown in the lower panel in Fig.~\ref{synth_10830} is far from perfect, but it shows that the profile is consistent with an emission component, possibly having the same ${ v_{\rm Dop}}$ and $\Delta  v_{\rm LOS}$ in He \textsc{i} 10830~\AA\,and He \textsc{i} D\textsubscript{3}. Results of this inversion are listed in Table~\ref{tabelsynth_10830}.

\begin{table}
\centering
\begin{scriptsize}
\caption{Retrieved parameters for two component inversion in Fig.~\ref{EBfits}. The upper, middle and lower row of parameters correspond to respectivily the upper, middle and lower row of fits in Fig.~\ref{EBfits}. \label{tabelebfits}}
\begin{tabular}{l r c r c r c}
\hline\hline
  & & & & & &  \\[-2mm]
  & EB1 & ranges & EB2 & ranges & EB3 & ranges \\[1mm]\hline
  & & & & & &\\[-1mm]
$\tau_1$              & 0.12  & 0 -- 1        & 0.99   & 0 -- 1        & 0.46  & 0 -- 1 \\[1mm]
${v_{\rm Dop,1}}\tablefootmark{1}$   & 12.70 & 3 -- 15       & 45.11  & 20 -- 70      & 25.47 & 10 -- 40 \\[1mm]
${v_{\rm LOS,1}}\tablefootmark{1}$ & -4.77 & -27.5 -- 2.5  & -23.52 & -37.5 -- -7.5 & 4.78  & -17.5 -- 22.5 \\[1mm]
$\beta_1$             & 3.43  & 3 -- 3.7      &  3.16  & 3 -- 3.25     & 3.61  & 3 -- 3.7 \\[1mm]
$\tau_2$              & 0.05  & 0 -- 1        & 1.46   & 0 --1         & 0.17  & 0 -- 1 \\[1mm]
${v_{\rm Dop,2}}\tablefootmark{1}$   & 8.66  & 5 -- 15       & 8.67   & 5 --11        & 8.05  & 5 -- 11 \\[1mm]
${v_{\rm LOS,2}}\tablefootmark{1}$ & 14.12 & -17.5 -- 42.5 & 3.42   & -17.5 -- 22.5 & 4.38  & -17.5 -- 22.5 \\[1mm]
$\beta_2$             & 1.96  & 0 -- 2        & 1.46   & 0 --2         & 0.94  & 0 -- 2 \\[1mm]   \hline
   & & & & & &\\[-1mm]
$\tau_1$              & 0.18  & 0 -- 1        & 0.95   & 0 -- 1        & 0.71  & 0 -- 1 \\[1mm]
${v_{\rm Dop,1}}\tablefootmark{1}$   & 14.88 & 5 -- 30       & 56.13  & 20 -- 70      & 18.88 & 10 -- 40 \\[1mm]
${v_{\rm LOS,1}}\tablefootmark{1}$ & -2.26 & -27.5 -- 2.5  & -12.15 & -37.5 -- -7.5 & -9.97 & -17.5 -- 22.5 \\[1mm]
$\beta_1$             & 3.71  & 3 -- 3.8      &  3.24  & 3 -- 3.25     & 3.53  & 3 -- 3.7 \\[1mm]
$\tau_2$              &       &               & 0.17   & 0 --1         & 0.17  & 0 -- 1 \\[1mm]
${v_{\rm Dop,2}}\tablefootmark{1}$   &       &               & 10.57  & 5 --11        & 9.66  & 5 -- 10 \\[1mm]
${v_{\rm LOS,2}}\tablefootmark{1}$ &       &               & 3.20   & -17.5 -- 22.5 & 3.95  & -17.5 -- 22.5 \\[1mm]
$\beta_2$             &       &               & 1.51   & 0 --2         & 0.58  & 0 -- 2 \\[1mm]   \hline
 & & & & & &\\[-1mm]
$\tau_1$              & 0.19  & 0 -- 1        & 0.71   & 0 -- 1        & 0.58  & 0 -- 1 \\[1mm]
${v_{\rm Dop,1}}\tablefootmark{1}$   & 12.18 & 5 -- 15       & 56.75  & 20 -- 80      & 24.35 & 10 -- 40 \\[1mm]
${v_{\rm LOS,1}}\tablefootmark{1}$ & 6.52  & -27.5 -- 12.5 & -10.46 & -37.5 -- -7.5 & -4.91 & -17.5 -- 22.5 \\[1mm]
$\beta_1$             & 1.71  & 0 -- 2        &  3.53  & 3 -- 3.7      & 3.60  & 3 -- 3.7 \\[1mm]
$\tau_2$              &       &               & 0.16   & 0 --1         & 0.19  & 0 -- 1 \\[1mm]
${v_{\rm Dop,2}}\tablefootmark{1}$   &       &               & 8.92   & 5 --11        & 7.55  & 7 -- 10 \\[1mm]
${v_{\rm LOS,2}}\tablefootmark{1}$ &       &               & 2.87   & -17.5 -- 22.5 & 5.08  & -17.5 -- 22.5 \\[1mm]
$\beta_2$             &       &               & 1.71   & 0 --2         & 0.95  & 0 -- 2 \\[1mm]   \hline
\end{tabular}
\newline
\tablefoottext{1}{Velocities are given in km~s$^{-1}$. Positive LOS velocities correspond to redshifts.}
\end{scriptsize}
\end{table}

\begin{table}
\centering
\caption{Retrieved parameters for the inversions in Fig.~\ref{synth_10830}. The upper and lower row of parameters correspond to respectivily the upper and lower row of fits in Fig.~\ref{synth_10830}. The parameters in bold have been kept constant during the inversion.\label{tabelsynth_10830}}
\begin{scriptsize}

\begin{tabular}{l r c}
\hline\hline
 & &   \\[-2mm]
 & EB2 & ranges \\[1mm]\hline
 & &   \\[-1mm]
$\tau$                & 0.18  & 0.1 -- 2       \\[1mm]
${v_{\rm Dop}}\tablefootmark{1}$     & 10.18 & 2 -- 15       \\[1mm]
${v_{\rm LOS}}\tablefootmark{1}$ & -6.69 & -10 -- 50  \\[1mm]\hline
   & & \\[-1mm]
$\tau_1$                & 0.20  & 0 -- 1        \\[1mm]
${v_{\rm Dop,1}}\tablefootmark{1}$     & \textbf{45.11}  &       \\[1mm]
${v_{\rm LOS,1}}\tablefootmark{1}$ & \textbf{-20.25} &  \\[1mm]
$\beta_1$               & 3.20  & 0 -- 2        \\[1mm]
$\tau_2$                & 1.20      &  0.5 -- 2.0             \\[1mm]
${v_{\rm Dop,2}}\tablefootmark{1}$     & \textbf{10.18}      &               \\[1mm]
${v_{\rm LOS,2}}\tablefootmark{1}$ & \textbf{-6.69}      &             \\[1mm]
$\beta_2$               & 0.88      &  0.1 -- 3.0             \\[1mm]   \hline
\end{tabular}
\newline
\tablefoottext{1}{Velocities are given in km~s$^{-1}$. Positive LOS velocities correspond to redshifts.}

\end{scriptsize}
\end{table}

\subsection{Conditions for detection of EB emission signatures in He \label{subs:cond}}
\begin{figure}[h]
\includegraphics[scale=1]{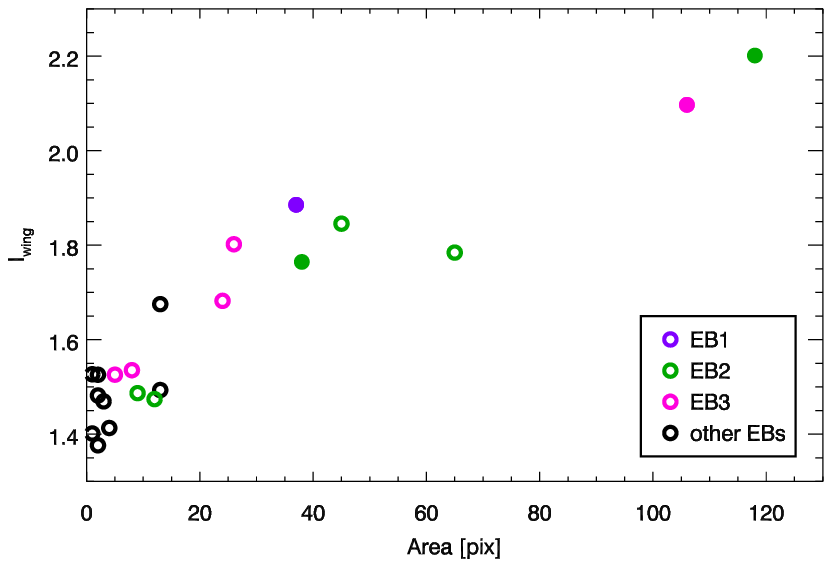}
\caption{EB1 is shown in purple, EB2 is shown in green and EB3 is shown in pink. The black circles correspond to the other EBs in our data, which do not show emission signals in He \textsc{i} D\textsubscript{3}. The different data points correspond to different time steps in which the bombs were detected. The filled circles are the detections showing emission in He \textsc{i} D\textsubscript{3}.  \label{area_int}}
\end{figure}


In our data, we have detected 21 EB events of which only 4 show emission signals in He \textsc{i} D\textsubscript{3} (the 4 events correspond to 3 EBs, of which EB2 has emission at 2 different time steps). We have tried to quantify conditions under which He \textsc{i} D\textsubscript{3} emission takes place. In Fig.\,\ref{area_int}, we show the $I_{\rm wing}$ of H$\beta$ of all the events (i.e. the intensity in the constructed wing image as described in Sec.\,\ref{subs:det}) as a function of their area. The area is the number of pixels satisfying the conditions for EB detection described in Sec.\,\ref{subs:det}. 

Fig.~\ref{area_int} demonstrates that EBs having He \textsc{i} D\textsubscript{3} emission are among the bigger and brighter events in H$\beta$. On the other hand, the detections of EBs in our data which do not show helium emission signals (black circles) are usually only a few pixels large, suggesting that we probably caught those EBs when they were just appearing or fading. We detected helium emission signal in three EBs which are very different: EB1 is relatively small and short-lived while EB2 is extremely large and long-lived, EB3 ranges between those two. This suggests that helium emission signals might be common in EBs if they are caught at the right time in their evolution.

Fig.~\ref{area_int} also shows that for some time steps, the bombs can be relatively bright in H$\beta$ while still not having He \textsc{i} D\textsubscript{3} emission and vice versa. We ruled out seeing effects by checking the Fried parameter $r_0$ for every event and found that the value of $r_0$ did not correlate with the detection of He \textsc{i} D\textsubscript{3} emission. We also compared the intensity of the bombs in SDO/AIA 1700 \AA\;and in the IRIS slit-jaw images. In all these wavelengths however, bright outliers are found which do not correspond to He \textsc{i} D\textsubscript{3} emission. Therefore, we cannot derive threshold values in H$\beta$, Si \textsc{iv}, Mg \textsc{ii} or AIA 1700 \AA\;for He \textsc{i} D\textsubscript{3} emission to take place, suggesting that timing, history and evolution of the event are crucial parameters. \citetads{2016A&A...590A.124R} suggested that EBs are heated up to $T\sim 20\,000$ K only in their onsets, but we detected emission signals at early as well as later stages of the evolution of the EBs, as can be seen in the light curve of EB2 (Fig.~\ref{LC2}). We suspect that effects of non-equilibrium time-dependent helium ionization are present \citepads{2014ApJ...784...30G}.

High-cadence data in He \textsc{i} D\textsubscript{3} would provide more clues on the conditions needed to create He \textsc{i} D\textsubscript{3} emission. SST/CRISP data could be useful for a study like this, however the limited spectral sampling and the lower spectral resolution of the instrument would make the detection of the emission signals more challenging than with SST/TRIPPEL. 

\section{Discussion}\label{sec:disc}
Using our analysis of the EB emission signals in helium, we argue on the cause of the emission. We should take into account the main observational features: the emission component is extremely broad and slightly blue-shifted, it can be either symmetrical or exhibit a blue asymmetry.

First of all, to generate He \textsc{i} D\textsubscript{3} emission, the neutral helium triplet levels need to be populated. Generally, in active regions, the triplet levels of neutral helium are populated by photoionization-recombination caused by EUV photons from the corona or transition region. It is clear that for the emission signals associated with the EBs, we can entirely ignore radiation coming from higher layers, which only influences the overlying absorption profile. We believe that the processes to ionize helium in EBs happen locally in these deep layers (upper photosphere - lower chromosphere). The levels should be populated by first ionizing helium through either locally generated EUV radiation or by collisions, and then recombining into the triplet states. Direct collisional excitation from the ground level is another possibility. In any case, high temperatures of the order of $T \ga 20\,000$ K are needed to achieve this (e.g. \citeads{2005A&A...432..699D}).

\begin{figure}
\includegraphics[scale=1]{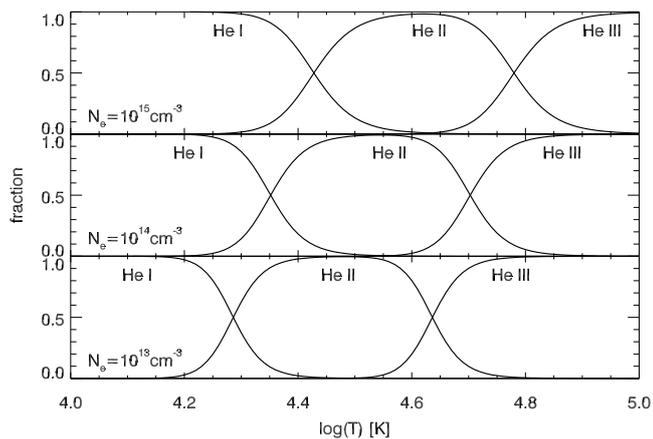}
\caption{Ionization diagrams for helium for different electron densities, calculated with the Saha equation. \label{CP_diag}}
\end{figure}

An upper limit for the temperature can be estimated by assuming that the broadening of the emission component is entirely caused by thermal Doppler motions. The Doppler motions of EB3 are ranging between 19 -- 25 km~s$^{-1}$, which yields $T\sim10^5$ K. For EB1 we have $v_{\rm Dop}$ between 12 -- 15 km~s$^{-1}$ corresponding to $T\sim 5\cdot10^4$ K. We choose to not apply this method to EB2 since the fitting results for these profiles are not satisfactory. The given temperature estimates serve as an upper limit because we have assumed that the micro-turbulence is zero, which is very unlikely. Using the Saha equation in LTE, we can make ionization diagrams as shown in Fig.~\ref{CP_diag}. LTE is an overly simplified method to model helium ionization in EBs, but it might provide a lower limit temperature estimate. From Fig~\ref{CP_diag}, we estimate that we need at least a temperature of $T\sim 2\cdot 10^4$ K to have a substantial fraction of He \textsc{ii}, for an electron density of $N_e=10^{14}\,\rm cm^{-3}$. Following \citetads{2016A&A...590A.124R} with a hydrogen density of $N_H\simeq N_e =10^{15}\,\rm cm^{-3}$, we obtain $T\sim 2.2\cdot 10^4$ K.

Are these high temperatures realistic? Temperatures of the order of $T\sim 10^5$ K have been proposed for IRIS bombs as well, in order to explain Si \textsc{iv} emission. In later papers, IBs have been shown to correspond to EBs in at least some cases (\citeads{2015ApJ...812...11V,2015ApJ...810...38K,2016ApJ...824...96T}). Our observations of IRIS profiles in Fig.~\ref{spec_withim10} also confirm that strong IRIS signatures can be associated with EBs. Our rough temperature estimate seems to agree with the general statement that EBs are probably much hotter than previously thought. \citetads{2015ApJ...812...11V} states that the violent heating occurs mainly in the top part of the EBs. Our observations also suggest that the highest temperatures are present only in a part of the bomb, and therefore the spatial extend of the bomb as seen in images in neutral helium is smaller than in H$\beta$.

\begin{figure}[h]
\includegraphics[scale=1]{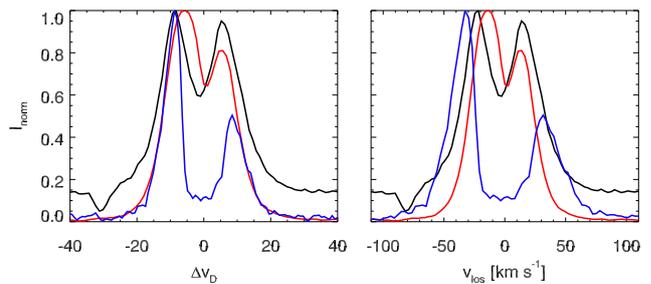}
\caption{Normalized intensity of IRIS lines Mg \textsc{ii} h (black), Si \textsc{iv} 1400 \AA\, (red) and C \textsc{ii} 1330 \AA\, (blue) overplotted on a Doppler width scale $\Delta v_{\rm D}$ and a LOS velocity scale $v_{\rm LOS}$. \label{vdop_scale}}
\end{figure}

To explain the extreme broadening of the helium emission component, we consider all possible broadening mechanisms. Instrumentational broadening is negligible but has been taken into account in the inversions, by convolving with the instrumentational profile of TRIPPEL. Collisions and radiative broadening only affect the damping. We have obtained reasonable fits with a small and constant damping parameter of $a=10^{-4}$, hence we do not suspect these effects to be dominating. Macroscopic velocity gradients can cause huge broadening, but as long as the emission component can be fitted with a constant slab, those effects are probably not dominating, as is the case for EB1 and EB3. EBs have often been associated with bi-directional jets but since we do not observe asymmetry in EB3, we suspect that the dominant part of the broadening is caused by thermal Doppler motions. This hypothesis is further supported by the similarity of the IRIS lines on a $\Delta v_{\rm D}$ scale instead of a $v_{\rm LOS}$ scale, see Fig. \ref{vdop_scale}. The C \textsc{ii} and Si \textsc{iv} lines overlap almost perfectly on the $\Delta v_{\rm D}$ scale. We are however careful with this interpretation, since we are unable to distinguish between thermal Doppler motions, micro-turbulence or turbulence on a macroscopic scale which would also cause symmetrical and broad profiles. 

The Si \textsc{iv} line in EB3 is rather symmetrical and not completely optically thin (peak intensity ratio of $\sim 1.5$). Therefore, we assign the central dip in the profile to a reversal in the source function, rather than the presence of bi-directional jets. We do still think that jets might be present, since most helium emission components are shifted to the blue with respect to the absorption component.

In the case of EB2, the strong asymmetry in the helium emission profiles (and also the H$\beta$ profiles) suggests that strong velocity gradients are present, possibly a bi-directional jet. However, in none of our bombs do we observe the red counterpart of the jet. One could suspect that the downward moving part of the jet dissipates quickly, and therefore we do not observe it. However, our observations are close to the limb so we cannot assign the blue part of the emission component to upwards moving material and the red part to downwards moving material in a straightforward way. 

So far we have assumed optically thin line formation for helium. However, if this is not the case, then the broadening of the emission component is influenced by the variation of the source function with height. The EB atmosphere must have a high-temperature bump located near the temperature minimum in the atmosphere. The location, shape and strength of this bump will influence the broadening and strength of the emission component. \textsc{Hazel} obviously does not take this effect into account since it uses a constant source function. 

Finally, we comment on the question why these EB helium emission signatures have been detected only now. \citetads{2015A&A...582A.104R} have observed EBs in He \textsc{i} 10830 \AA, but only reported co-spatial absorption. They propose to use the line to reject false EB identifications by showing emission profiles. Our detections strongly discourage this idea. There is a variety of reasons for why \citetads{2015A&A...582A.104R} have not been able to detect the helium emission signals. First of all, they give a value of 1\arcsec\;as spatial resolution for their observations. The area of the emission signatures in helium is of the order of 1\arcsec\;or smaller, as shown in the raster scan images of He \textsc{i} D\textsubscript{3} and He \textsc{i} 10830 \AA\;in Figs.~\ref{spec_withim}, \ref{spec_withim5} and \ref{spec_withim10}. The helium emission signatures have low contrast so seeing effects easily reduce the signal to non-detectable limits, especially when using long integration times (since \citetads{2015A&A...582A.104R} were doing polarimetric measurments, they used an integration time of 6 s per slit position, as opposed to our exposure times of 100 ms which we used for non-polarimetric observations). Also, as discussed before, the emission signals are much harder to detect in He \textsc{i} 10830 \AA\;than in He \textsc{i} D\textsubscript{3}. A last possibility for non-detection is that the helium emission signals were simply not present. In Sec.~\ref{subs:cond}, we have shown that EBs do not always have helium emission signal even when the H$\beta$ emission is strong.

On the same topic of detection, we should consider the fact that our observations are taken close to the limb and that helium triplet lines are known to be limb-brightened (e.g. \citeads{1975ApJ...199L..63Z}). Closer to disk center, it might be possible that the helium emission signals would not exceed the continuum intensity. 

\section{Summary and Conclusions \label{sec:concl}}
In our data, we have detected 21 EBs of which four (i.e. EB1, EB2a, EB2b and EB3) show emission signals in He \textsc{i} D\textsubscript{3} and He \textsc{i} 10830 \AA. The EB profiles of these lines show two components: a very broad and slightly blue-shifted emission profile and a more conventional looking absorption profile. The emission component is associated with the EB, formed in deeper layers (upper photosphere - lower chromosphere), while the absorption component is probably formed in the upper chromosphere where the neutral helium triplet lines are usually formed. The signals are harder to detect in He \textsc{i} 10830~\AA\,because the overlying absorption profile is stronger, and the wings of the line have strong blends.

Inversions of the He \textsc{i} D\textsubscript{3} profiles with \textsc{Hazel} reveal that the emission component of EB3 is symmetrical, while the emission component of EB2 exhibits strong asymmetry in the blue. 

We suspect the extreme broadening to be mostly caused by thermal Doppler motions, supported by coincidence of the Si \textsc{iv} and C \textsc{ii} line on a $\Delta v_{\rm D}$-scale. We assign the peak reversal of Si \textsc{iv} to opacity effects. The strong blue asymmetry observed in the helium emission component of EB2 is interpreted as caused by strong velocity gradients, possibly bi-directional jets.

A rough order of magnitude estimate places our observed EBs at temperatures between $T\sim 2\cdot 10^4 - 10^5$. The analysis of our observations has now put us in an ideal starting position for doing more detailed modelling of the bombs. Using the extra constraints of neutral helium triplet emission and the signatures in IRIS lines, will most likely lead to more reliable estimates of the temperatures and formation height of EBs. 

Another natural continuation would be to observe EBs in all Stokes parameters at higher cadence (with e.g. SST/CRISP). A study like this might reveal clues on the physics causing emission in helium and possibly on magnetic conditions in EBs.

\begin{acknowledgements}
We thank Guus Sliepen for implementing the possibility to use scanning with the correlation tracker at the SST. We thank Dan Kiselman and Pit S\"utterlin for assistance with the observations. We thank Phil Judge, G\"oran Scharmer and Dan Kiselman for comments which improved the content of the paper. We also thank the anonymous referee for suggestions which have improved the manuscript. We made use of routines by Rob Rutten to align SDO data with SST data. 

TL, JJ and JdlCR acknowledge financial support from the
\textsc{Chromobs} project funded by the Knut and Alice Wallenberg
Foundation. JdlCR is supported by grants from the Swedish Research Council and the Swedish National Space Board.
AAR acknowledges financial support by the Spanish Ministry of Economy and Competitiveness 
through projects AYA2014-60476-P and Consolider-Ingenio 2010 CSD2009-00038. AAR also acknowledges financial support
through the Ram\'on y Cajal fellowships.

The Swedish 1-m Solar Telescope is operated on the island of La Palma
by the Institute for Solar Physics of Stockholm University in the
Spanish Observatorio del Roque de los Muchachos of the Instituto de
Astrof\'isica de Canarias. 

This research has made use of NASA's
Astrophysics Data System Bibliographic Services.
\end{acknowledgements}

\bibliography{references}

\begin{thebibliography}{51}
\expandafter\ifx\csname natexlab\endcsname\relax\def\natexlab#1{#1}\fi

\bibitem[{{Andretta} {et~al.}(2008){Andretta}, {Mauas}, {Falchi}, \&
  {Teriaca}}]{2008ApJ...681..650A}
{Andretta}, V., {Mauas}, P.~J.~D., {Falchi}, A., \& {Teriaca}, L. 2008, \apj,
  681, 650

\bibitem[{{Asensio Ramos} {et~al.}(2008){Asensio Ramos}, {Trujillo Bueno}, \&
  {Landi Degl'Innocenti}}]{2008ApJ...683..542A}
{Asensio Ramos}, A., {Trujillo Bueno}, J., \& {Landi Degl'Innocenti}, E. 2008,
  \apj, 683, 542

\bibitem[{{Bello Gonz{\'a}lez} {et~al.}(2013){Bello Gonz{\'a}lez}, {Danilovic},
  \& {Kneer}}]{2013A&A...557A.102B}
{Bello Gonz{\'a}lez}, N., {Danilovic}, S., \& {Kneer}, F. 2013, \aap, 557, A102

\bibitem[{{Berlicki} \& {Heinzel}(2014)}]{2014A&A...567A.110B}
{Berlicki}, A. \& {Heinzel}, P. 2014, \aap, 567, A110

\bibitem[{{Bruzek}(1972)}]{1972SoPh...26...94B}
{Bruzek}, A. 1972, \solphys, 26, 94

\bibitem[{{Centeno} {et~al.}(2008){Centeno}, {Trujillo Bueno}, {Uitenbroek}, \&
  {Collados}}]{2008ApJ...677..742C}
{Centeno}, R., {Trujillo Bueno}, J., {Uitenbroek}, H., \& {Collados}, M. 2008,
  \apj, 677, 742

\bibitem[{{Dara} {et~al.}(1997){Dara}, {Alissandrakis}, {Zachariadis}, \&
  {Georgakilas}}]{1997A&A...322..653D}
{Dara}, H.~C., {Alissandrakis}, C.~E., {Zachariadis}, T.~G., \& {Georgakilas},
  A.~A. 1997, \aap, 322, 653

\bibitem[{{De Pontieu} {et~al.}(2014){De Pontieu}, {Title}, {Lemen}, {Kushner},
  {Akin}, {Allard}, {Berger}, {Boerner}, {Cheung}, {Chou}, {Drake}, {Duncan},
  {Freeland}, {Heyman}, {Hoffman}, {Hurlburt}, {Lindgren}, {Mathur}, {Rehse},
  {Sabolish}, {Seguin}, {Schrijver}, {Tarbell}, {W{\"u}lser}, {Wolfson},
  {Yanari}, {Mudge}, {Nguyen-Phuc}, {Timmons}, {van Bezooijen}, {Weingrod},
  {Brookner}, {Butcher}, {Dougherty}, {Eder}, {Knagenhjelm}, {Larsen},
  {Mansir}, {Phan}, {Boyle}, {Cheimets}, {DeLuca}, {Golub}, {Gates}, {Hertz},
  {McKillop}, {Park}, {Perry}, {Podgorski}, {Reeves}, {Saar}, {Testa}, {Tian},
  {Weber}, {Dunn}, {Eccles}, {Jaeggli}, {Kankelborg}, {Mashburn}, {Pust},
  {Springer}, {Carvalho}, {Kleint}, {Marmie}, {Mazmanian}, {Pereira}, {Sawyer},
  {Strong}, {Worden}, {Carlsson}, {Hansteen}, {Leenaarts}, {Wiesmann},
  {Aloise}, {Chu}, {Bush}, {Scherrer}, {Brekke}, {Martinez-Sykora}, {Lites},
  {McIntosh}, {Uitenbroek}, {Okamoto}, {Gummin}, {Auker}, {Jerram}, {Pool}, \&
  {Waltham}}]{2014SoPh..289.2733D}
{De Pontieu}, B., {Title}, A.~M., {Lemen}, J.~R., {et~al.} 2014, \solphys, 289,
  2733

\bibitem[{{Ding} {et~al.}(2005){Ding}, {Li}, \& {Fang}}]{2005A&A...432..699D}
{Ding}, M.~D., {Li}, H., \& {Fang}, C. 2005, \aap, 432, 699

\bibitem[{{Ellerman}(1917)}]{1917ApJ....46..298E}
{Ellerman}, F. 1917, \apj, 46, 298

\bibitem[{{Fang} {et~al.}(2006){Fang}, {Tang}, {Xu}, {Ding}, \&
  {Chen}}]{2006ApJ...643.1325F}
{Fang}, C., {Tang}, Y.~H., {Xu}, Z., {Ding}, M.~D., \& {Chen}, P.~F. 2006,
  \apj, 643, 1325

\bibitem[{{Georgoulis} {et~al.}(2002){Georgoulis}, {Rust}, {Bernasconi}, \&
  {Schmieder}}]{2002ApJ...575..506G}
{Georgoulis}, M.~K., {Rust}, D.~M., {Bernasconi}, P.~N., \& {Schmieder}, B.
  2002, \apj, 575, 506

\bibitem[{{Goldberg}(1939)}]{1939ApJ....89..673G}
{Goldberg}, L. 1939, \apj, 89, 673

\bibitem[{{Golding} {et~al.}(2014){Golding}, {Carlsson}, \&
  {Leenaarts}}]{2014ApJ...784...30G}
{Golding}, T.~P., {Carlsson}, M., \& {Leenaarts}, J. 2014, \apj, 784, 30

\bibitem[{{Hashimoto} {et~al.}(2010){Hashimoto}, {Kitai}, {Ichimoto}, {Ueno},
  {Nagata}, {Ishii}, {Hagino}, {Komori}, {Nishida}, {Matsumoto}, {Otsuji},
  {Nakamura}, {Kawate}, {Watanabe}, \& {Shibata}}]{2010PASJ...62..879H}
{Hashimoto}, Y., {Kitai}, R., {Ichimoto}, K., {et~al.} 2010, \pasj, 62, 879

\bibitem[{{Hong} {et~al.}(2014){Hong}, {Ding}, {Li}, {Fang}, \&
  {Cao}}]{2014ApJ...792...13H}
{Hong}, J., {Ding}, M.~D., {Li}, Y., {Fang}, C., \& {Cao}, W. 2014, \apj, 792,
  13

\bibitem[{{Judge}(2015)}]{2015ApJ...808..116J}
{Judge}, P.~G. 2015, \apj, 808, 116

\bibitem[{{Kim} {et~al.}(2015){Kim}, {Yurchyshyn}, {Bong}, {Cho}, {Cho}, {Lee},
  {Lim}, {Park}, {Yang}, {Ahn}, {Goode}, \& {Jang}}]{2015ApJ...810...38K}
{Kim}, Y.-H., {Yurchyshyn}, V., {Bong}, S.-C., {et~al.} 2015, \apj, 810, 38

\bibitem[{{Kiselman} {et~al.}(2011){Kiselman}, {Pereira}, {Gustafsson},
  {Asplund}, {Mel{\'e}ndez}, \& {Langhans}}]{2011A&A...535A..14K}
{Kiselman}, D., {Pereira}, T.~M.~D., {Gustafsson}, B., {et~al.} 2011, \aap,
  535, A14

\bibitem[{{Kitai}(1983)}]{1983SoPh...87..135K}
{Kitai}, R. 1983, \solphys, 87, 135

\bibitem[{Kramida {et~al.}(2015)Kramida, {Yu.~Ralchenko}, Reader, \& {and NIST
  ASD Team}}]{NIST_ASD}
Kramida, A., {Yu.~Ralchenko}, Reader, J., \& {and NIST ASD Team}. 2015, {NIST
  Atomic Spectra Database (ver. 5.3), [Online]. Available:
  {\tt{http://physics.nist.gov/asd}} [2016, September 1]. National Institute of
  Standards and Technology, Gaithersburg, MD.}

\bibitem[{{Lagg} {et~al.}(2004){Lagg}, {Woch}, {Krupp}, \&
  {Solanki}}]{2004A&A...414.1109L}
{Lagg}, A., {Woch}, J., {Krupp}, N., \& {Solanki}, S.~K. 2004, \aap, 414, 1109

\bibitem[{{Leenaarts} {et~al.}(2016){Leenaarts}, {Golding}, {Carlsson},
  {Libbrecht}, \& {Joshi}}]{2016arXiv160800838L}
{Leenaarts}, J., {Golding}, T., {Carlsson}, M., {Libbrecht}, T., \& {Joshi}, J.
  2016, ArXiv e-prints [\eprint[arXiv]{1608.00838}]

\bibitem[{{Lemen} {et~al.}(2012){Lemen}, {Title}, {Akin}, {Boerner}, {Chou},
  {Drake}, {Duncan}, {Edwards}, {Friedlaender}, {Heyman}, {Hurlburt}, {Katz},
  {Kushner}, {Levay}, {Lindgren}, {Mathur}, {McFeaters}, {Mitchell}, {Rehse},
  {Schrijver}, {Springer}, {Stern}, {Tarbell}, {Wuelser}, {Wolfson}, {Yanari},
  {Bookbinder}, {Cheimets}, {Caldwell}, {Deluca}, {Gates}, {Golub}, {Park},
  {Podgorski}, {Bush}, {Scherrer}, {Gummin}, {Smith}, {Auker}, {Jerram},
  {Pool}, {Soufli}, {Windt}, {Beardsley}, {Clapp}, {Lang}, \&
  {Waltham}}]{2012SoPh..275...17L}
{Lemen}, J.~R., {Title}, A.~M., {Akin}, D.~J., {et~al.} 2012, \solphys, 275, 17

\bibitem[{{Li} {et~al.}(2015){Li}, {Fang}, {Guo}, {Chen}, {Xu}, \&
  {Cao}}]{2015RAA....15.1513L}
{Li}, Z., {Fang}, C., {Guo}, Y., {et~al.} 2015, Research in Astronomy and
  Astrophysics, 15, 1513

\bibitem[{{Mauas} {et~al.}(2005){Mauas}, {Andretta}, {Falchi}, {Falciani},
  {Teriaca}, \& {Cauzzi}}]{2005ApJ...619..604M}
{Mauas}, P.~J.~D., {Andretta}, V., {Falchi}, A., {et~al.} 2005, \apj, 619, 604

\bibitem[{{Mitchell}(1909)}]{1909ApJ....30...75M}
{Mitchell}, W.~M. 1909, \apj, 30, 75

\bibitem[{{Moore} {et~al.}(1966){Moore}, {Minnaert}, \& {Houtgast}}]{moore}
{Moore}, C.~E., {Minnaert}, M. G.~J., \& {Houtgast}, J. 1966, The Solar
  Spectrum 2935A to 8770A Second Revision of Rowland's Preliminary Table of
  Solar Spectrum Wavelelengths, ed. N.~B. of~Standards (United States
  Department of Commerce)

\bibitem[{{Neckel} \& {Labs}(1984)}]{1984SoPh...90..205N}
{Neckel}, H. \& {Labs}, D. 1984, \solphys, 90, 205

\bibitem[{{Nelson} {et~al.}(2013){Nelson}, {Shelyag}, {Mathioudakis}, {Doyle},
  {Madjarska}, {Uitenbroek}, \& {Erd{\'e}lyi}}]{2013ApJ...779..125N}
{Nelson}, C.~J., {Shelyag}, S., {Mathioudakis}, M., {et~al.} 2013, \apj, 779,
  125

\bibitem[{{Pariat} {et~al.}(2004){Pariat}, {Aulanier}, {Schmieder},
  {Georgoulis}, {Rust}, \& {Bernasconi}}]{2004ApJ...614.1099P}
{Pariat}, E., {Aulanier}, G., {Schmieder}, B., {et~al.} 2004, \apj, 614, 1099

\bibitem[{{Pariat} {et~al.}(2007){Pariat}, {Schmieder}, {Berlicki}, {Deng},
  {Mein}, {L{\'o}pez Ariste}, \& {Wang}}]{2007A&A...473..279P}
{Pariat}, E., {Schmieder}, B., {Berlicki}, A., {et~al.} 2007, \aap, 473, 279

\bibitem[{{Pereira} {et~al.}(2015){Pereira}, {Carlsson}, {De Pontieu}, \&
  {Hansteen}}]{2015ApJ...806...14P}
{Pereira}, T.~M.~D., {Carlsson}, M., {De Pontieu}, B., \& {Hansteen}, V. 2015,
  \apj, 806, 14

\bibitem[{{Pereira} {et~al.}(2009){Pereira}, {Kiselman}, \&
  {Asplund}}]{2009A&A...507..417P}
{Pereira}, T.~M.~D., {Kiselman}, D., \& {Asplund}, M. 2009, \aap, 507, 417

\bibitem[{{Peter} {et~al.}(2014){Peter}, {Tian}, {Curdt}, {Schmit}, {Innes},
  {De Pontieu}, {Lemen}, {Title}, {Boerner}, {Hurlburt}, {Tarbell}, {Wuelser},
  {Mart{\'{\i}}nez-Sykora}, {Kleint}, {Golub}, {McKillop}, {Reeves}, {Saar},
  {Testa}, {Kankelborg}, {Jaeggli}, {Carlsson}, \&
  {Hansteen}}]{2014Sci...346C.315P}
{Peter}, H., {Tian}, H., {Curdt}, W., {et~al.} 2014, Science, 346, 1255726

\bibitem[{{Qiu} {et~al.}(2000){Qiu}, {Ding}, {Wang}, {Denker}, \&
  {Goode}}]{2000ApJ...544L.157Q}
{Qiu}, J., {Ding}, M.~D., {Wang}, H., {Denker}, C., \& {Goode}, P.~R. 2000,
  \apjl, 544, L157

\bibitem[{{Reid} {et~al.}(2016){Reid}, {Mathioudakis}, {Doyle}, {Scullion},
  {Nelson}, {Henriques}, \& {Ray}}]{2016ApJ...823..110R}
{Reid}, A., {Mathioudakis}, M., {Doyle}, J.~G., {et~al.} 2016, \apj, 823, 110

\bibitem[{{Rezaei} \& {Beck}(2015)}]{2015A&A...582A.104R}
{Rezaei}, R. \& {Beck}, C. 2015, \aap, 582, A104

\bibitem[{{Rouppe van der Voort} {et~al.}(2016){Rouppe van der Voort},
  {Rutten}, \& {Vissers}}]{2016arXiv160603675R}
{Rouppe van der Voort}, L.~H.~M., {Rutten}, R.~J., \& {Vissers}, G.~J.~M. 2016,
  ArXiv e-prints [\eprint[arXiv]{1606.03675}]

\bibitem[{{Rutten}(2016)}]{2016A&A...590A.124R}
{Rutten}, R.~J. 2016, \aap, 590, A124

\bibitem[{{Rutten} {et~al.}(2015){Rutten}, {Rouppe van der Voort}, \&
  {Vissers}}]{2015ApJ...808..133R}
{Rutten}, R.~J., {Rouppe van der Voort}, L.~H.~M., \& {Vissers}, G.~J.~M. 2015,
  \apj, 808, 133

\bibitem[{{Rutten} {et~al.}(2013){Rutten}, {Vissers}, {Rouppe van der Voort},
  {S{\"u}tterlin}, \& {Vitas}}]{2013JPhCS.440a2007R}
{Rutten}, R.~J., {Vissers}, G.~J.~M., {Rouppe van der Voort}, L.~H.~M.,
  {S{\"u}tterlin}, P., \& {Vitas}, N. 2013, Journal of Physics Conference
  Series, 440, 012007

\bibitem[{{Scharmer} {et~al.}(2003){Scharmer}, {Bjelksjo}, {Korhonen},
  {Lindberg}, \& {Petterson}}]{2003SPIE.4853..341S}
{Scharmer}, G.~B., {Bjelksjo}, K., {Korhonen}, T.~K., {Lindberg}, B., \&
  {Petterson}, B. 2003, in Society of Photo-Optical Instrumentation Engineers
  (SPIE) Conference Series, Vol. 4853, Innovative Telescopes and
  Instrumentation for Solar Astrophysics, ed. S.~L. {Keil} \& S.~V. {Avakyan},
  341--350

\bibitem[{{Schou} {et~al.}(2012){Schou}, {Scherrer}, {Bush}, {Wachter},
  {Couvidat}, {Rabello-Soares}, {Bogart}, {Hoeksema}, {Liu}, {Duvall}, {Akin},
  {Allard}, {Miles}, {Rairden}, {Shine}, {Tarbell}, {Title}, {Wolfson},
  {Elmore}, {Norton}, \& {Tomczyk}}]{2012SoPh..275..229S}
{Schou}, J., {Scherrer}, P.~H., {Bush}, R.~I., {et~al.} 2012, \solphys, 275,
  229

\bibitem[{{Socas-Navarro} {et~al.}(2006){Socas-Navarro}, {Mart{\'{\i}}nez
  Pillet}, {Elmore}, {Pietarila}, {Lites}, \& {Manso
  Sainz}}]{2006SoPh..235...75S}
{Socas-Navarro}, H., {Mart{\'{\i}}nez Pillet}, V., {Elmore}, D., {et~al.} 2006,
  \solphys, 235, 75

\bibitem[{{Tian} {et~al.}(2016){Tian}, {Xu}, {He}, \&
  {Madsen}}]{2016ApJ...824...96T}
{Tian}, H., {Xu}, Z., {He}, J., \& {Madsen}, C. 2016, \apj, 824, 96

\bibitem[{{Vissers} {et~al.}(2013){Vissers}, {Rouppe van der Voort}, \&
  {Rutten}}]{2013ApJ...774...32V}
{Vissers}, G.~J.~M., {Rouppe van der Voort}, L.~H.~M., \& {Rutten}, R.~J. 2013,
  \apj, 774, 32

\bibitem[{{Vissers} {et~al.}(2015){Vissers}, {Rouppe van der Voort}, {Rutten},
  {Carlsson}, \& {De Pontieu}}]{2015ApJ...812...11V}
{Vissers}, G.~J.~M., {Rouppe van der Voort}, L.~H.~M., {Rutten}, R.~J.,
  {Carlsson}, M., \& {De Pontieu}, B. 2015, \apj, 812, 11

\bibitem[{{Watanabe} {et~al.}(2008){Watanabe}, {Kitai}, {Okamoto}, {Nishida},
  {Kiyohara}, {Ueno}, {Hagino}, {Ishii}, \& {Shibata}}]{2008ApJ...684..736W}
{Watanabe}, H., {Kitai}, R., {Okamoto}, K., {et~al.} 2008, \apj, 684, 736

\bibitem[{{Watanabe} {et~al.}(2011){Watanabe}, {Vissers}, {Kitai}, {Rouppe van
  der Voort}, \& {Rutten}}]{2011ApJ...736...71W}
{Watanabe}, H., {Vissers}, G., {Kitai}, R., {Rouppe van der Voort}, L., \&
  {Rutten}, R.~J. 2011, \apj, 736, 71

\bibitem[{{Zirin}(1975)}]{1975ApJ...199L..63Z}
{Zirin}, H. 1975, \apjl, 199, L63

\end{thebibliography}

\end{document}